\documentclass[letterpaper,twocolumn,10pt]{article}
\pdfoutput=1
\usepackage{usenix, graphicx, xspace, subfigure,outlines,url,amsthm,grffile,hyperref,enumerate,listings}
\usepackage[compact]{titlesec}
\usepackage[normalem]{ulem}
\usepackage{paralist}
\usepackage{amsmath}
\usepackage{xcolor}
\usepackage{changebar}
\usepackage[footnotesize]{caption}
\usepackage{microtype}
\usepackage{flushend} 
\usepackage{cite}
\usepackage{times}
\usepackage{enumitem}
\setitemize{itemsep=1pt,topsep=1pt,parsep=1pt,partopsep=1pt}
\newcommand{\bi}{\begin{itemize}}
\newcommand{\ei}{\end{itemize}}

\newcommand{\eg}{{\it e.g.,}\xspace}
\newcommand{\ie}{{\it i.e.,}\xspace}
\newcommand\eat[1]{}
\newcommand\paragraphb[1]{\noindent{\bf #1}}

\newcommand{\allnotes}[1]{}
\renewcommand{\allnotes}[1]{\textit{#1}}
\newcommand{\fixme}[1]{\allnotes{\bf\textcolor{red}{[#1]}}}

\newcommand{\noteori}[1]{\allnotes{\textcolor{green}{[Ori: #1]}}}
\newcommand{\notemooly}[1]{\allnotes{\textcolor{purple}{[Mooly: #1]}}}

\newcommand{\notepanda}[1]{\allnotes{\textcolor{cyan}{[Panda: #1]}}}

\def\squarebox#1{\hbox to #1{\hfill\vbox to #1{\vfill}}}

\newcommand{\subsecref}[1]{\S\ref{subsec:#1}}

\usepackage{framed}
\usepackage{titling}

\colorlet{shadecolor}{gray!25}   
{\normalsize \endMakeFramed}
\title{Verifying Isolation Properties in the Presence of Middleboxes}
\author{
\rm{Aurojit Panda$^\ddag$, Ori Lahav$^\natural$, Katerina Argyraki$^\dag$, Mooly Sagiv$^\diamond$, Scott Shenker$^{\ddag\spadesuit}$}\\
$^\ddag$ UC Berkeley, $^\natural$MPI SWS, $^\dag$ EPFL, $^\diamond$ TAU, $^\spadesuit$ICSI\vspace{-2.4in}
}
\date{}
\begin{document}
\setlength{\abovedisplayskip}{0.2ex}
\setlength{\belowdisplayskip}{0.2ex}
\setlength{\abovedisplayshortskip}{0.2ex}
\setlength{\belowdisplayshortskip}{0.2ex}

\vspace{-1.5in}
\maketitle
\vspace{-1.8in}
\vskip -1.5em
\begin{abstract}
Great progress has been made recently in verifying the correctness of router forwarding tables \cite{mai2011debugging,khurshid2012veriflow,kazemian2012header,kazemian2013real}. However, these approaches do not work
for networks containing middleboxes such as caches and firewalls whose forwarding behavior depends on previously observed traffic.
We explore how to verify isolation properties in networks that include such ``dynamic datapath" elements using model checking.
Our work leverages recent advances in SMT solvers, and the main challenge lies
in scaling the approach to handle large and complicated networks.
While the straightforward application of model checking to this problem can only handle very small networks (if at all), our approach can verify simple realistic invariants on networks containing 
30,000 middleboxes in a few minutes.
\end{abstract}

\section{Introduction}
\label{sec:introduction}
\eat{
- Verification in networking normally taken to imply checks for loop freedom and no blackholes. In this paper using verification to talk about isolation (or use some other word).\\

}

Perhaps lulled into a sense of complacency because of the Internet's best-effort delivery model, which makes no explicit promises about network behavior, networking has long relied on ad hoc configuration and a ``we'll fix it when it breaks" operational attitude. However, as networking matures as a field, and institutions increasingly rely on networks to provide isolation and other behavioral guarantees, there is growing interest in developing rigorous verification tools that can ensure the correctness of network configurations. The first works along this line 
-- Anteater \cite{mai2011debugging}, Veriflow \cite{khurshid2012veriflow}, and HSA \cite{kazemian2012header,kazemian2013real} -- provide highly efficient (in fact, near real-time) checking of connectivity (and, conversely isolation) properties and detect anomalies such as loops and black holes. This represents a massive and invaluable step forward for networking.

These verification tools assume that the forwarding behavior is set by the control plane, and not altered by the traffic, so verification needs to be invoked only when the control plane alters routing entries. This approach is entirely sufficient for networks of routers, which is obviously an important use case. However, modern networks contain more than routers. 

Most networks contain switches whose learning behavior renders their forwarding behavior dependent on the traffic they have seen. More generally, most networks also contain middleboxes, and middleboxes often have forwarding behavior that depends on the observed traffic. For instance, content caches forward differently based on whether the desired content is found locally, and firewalls often rely on outbound ``hole-punching" to allow flows to enter an enterprise network. We will refer to network elements whose forwarding behavior can be affected by datapath activity as having a ``dynamic datapath", and additional examples of such elements include WAN optimizers, deep-packet-inspection boxes, and load balancers.

While classical networking often treats middleboxes as an unfortunate and rare occurrence in networks, the reality is that middleboxes have become the most viable way of incrementally deploying new network functionality. Operators have turned to middleboxes to such a great extent that a recent study \cite{sherry2012making} of over one hundred enterprise networks revealed that these networks are roughly equally divided between routers, switches and middleboxes. Thus, roughly two-thirds of the forwarding boxes in enterprise networks have dynamic datapaths, and do not abide by the models used in the recently developed network verification tools. Moreover, the rise of Network Function Virtualization (NFV)~\cite{nfv}, in which physical middleboxes are replaced by their virtual counterparts, makes it easier to deploy additional middleboxes without changes in the physical infrastructure. Thus, we must reconcile ourselves to the fact that many networks will have substantial numbers of elements with dynamic datapaths.

Not only are middleboxes prevalent, but they are often responsible for network problems. In December 2012 a misconfiguration in Google's load balancers resulted in a several minute outage for GMail and other Google Services~\cite{googleoutage}. A recent two year study~\cite{potharaju2013demystifying} of a provider found that middleboxes played a role in 43\% of their failure incidents, and between 4 and 15\% of these failures were the result of middlebox misconfiguration. Thus, middleboxes are a significant cause of network problems, and we have no verification tools that can help.

The goal of this paper is to extend the notion of verification to networks containing dynamic datapaths, so that we can check if invariants such 
as connectivity or isolation hold. Our basic approach is simple: we treat each
dynamic datapath element as a ``subroutine" and the network as a whole as a program. The routers and switches provide the glue
that connects these procedures.  The specified invariants imply constraints on dataflow within this program. We use symbolic model checking to determine
if the specified invariants hold. As described so far, this is a straightforward application of standard programming language
techniques to networks. However, na\"ively applied, this approach would fail to scale: middlebox code is extremely complicated,
and checking even simple invariants in modest-sized networks would be intractable. Thus, the
bulk of this paper, and the focus of our contribution, is about how to scale this approach to large networks. 

Our efforts to scale to large networks involves four different aspects:

{\bf 1. Limited invariants}: Rather than deal with an arbitrary set of invariants, we focus on two specific categories.  The first category of invariants involve the packet processing requirements (as defined by the operator) for various classes of packets; these requirements are specified as a set of middleboxes (more generally as a DAG of middleboxes) packets should flow through; we call these {\it pipeline} invariants. The second category of invariants concern the overall behavior of the network, and here we only consider invariants that address reachability and isolation between hosts (at the packet and content level). These pipeline and isolation invariants play distinct roles in the network, and their verification is done quite differently.

{\bf 2. Simple high-level middlebox models}: A standard approach to model checking middleboxes would use their full implementation. This is infeasible for two reasons: (i) we do not have access to middlebox code, and (ii) model checking even one such box for even the simplest invariants would be difficult. Instead, we consider simplified models (as we discuss later, these reduced models capture only the dependence on the packet header). These models can typically be derived from a general description of the middlebox's behavior and can be represented as a state machine that can be easily analyzed. 

{\bf 3. Modularized network models}: Networks contain elements with static datapaths and dynamic datapaths. Rather than consider them all within the model checking framework, which would overburden a system already having trouble scaling, we treat the two separately: it is the job of the static datapath elements to satisfy the pipeline invariants (that is, to carry packets through the appropriate set of middleboxes), which we can analyze using existing verification tools; and it is the job of the processing pipeline to enforce isolation invariants, and that is where we focus our attention.  Thus, our resulting system is a hybrid of current static-datapath verification tools and our newly-proposed tool for dynamic datapaths. 

{\bf 4. Special class of network enforcement}: Certain network designs allow invariants to be verified by checking only a portion of the network. We show that these designs
carry no additional overhead but allow operators to quickly verify their policies; the use of such designs is key to scaling out verification to large networks.
\bigskip

These four steps lead us to a system that can verify realistic invariants on very large networks; as an example, we can verify a set of isolation invariants on a network containing 30,000 middleboxes in $2$ to $5$ minutes.

In the next section, we discuss all four of these steps more formally, and then in \S\ref{sec:system} we provide an overview of the system we built that incorporates these ideas. We provide a theoretical analysis of the tractability of our approach in \S\ref{sec:tractability} and provide performance numbers from our operational system in \S\ref{sec:eval}. We conclude in \S\ref{sec:related} and \S\ref{sec:conclusion} with a discussion of related work and a brief summary.

\section{Background}
\label{sec:modeling}
\eat{
In this section we define the verification problem addressed by this paper.
\subsecref{modeling:invariants} defines the class of desired network properties.
Our system can also prove other properties.
However, focusing on specific properties allows us to abstract many features of middlebox which are hard for automatic reasoning.
\subsecref{modeling-mboxes} formalizes the effect of middleboxes on packets and defines natural restrictions on their behaviors.
Notice that we are not interested in proving that the middleboxes correctly implement their functionality.
Instead, we implemented a system which verifies that a network of middleboxes satisfies the desired properties using an SMT solver~\cite{de2008z3}.
The effect of each of the individual middlebox is specified using a logical formula which captures all possible behaviors of the middlebox relevant to
the desired network properties.

\subsecref{pipeline} defines an additional property of middleboxes which permits connecting several middleboxes in a pipeline.
This is explored in \subsecref{modeling:stronger} by defining the notion of RONO subnetworks ---
pipelines of middleboxes whose behavior does not depend on the effect of other middleboxes.
}
We begin by defining the verification problem addressed by this paper and describing the simplification steps that allow
our system (described in \S\ref{sec:system}) to scale. First we present the specific invariants we analyze (\S\ref{subsec:modeling:invariants}).
Next we show that by focusing on these specific invariants and using some natural restriction on middlebox behaviors (\S\ref{subsec:modeling-mboxes}) we can
greatly simplify automated reasoning and verification for these middleboxes. Next in \subsecref{pipeline} we show how multiple middlebox models can be combined
so we can reason about a network and finally (\S\ref{subsec:modeling:stronger}) we find some additional conditions that allow us to verify network wide properties
by operating on individual pipelines instead of the entire network.

Note we do not attempt to verify that middlebox implementations are correct (\ie obey the given model). However, we do discuss how one can {\it enforce} that middleboxes obey the abstract model by simulating the state-machine that models its intended behavior. But this enforcement is merely a small aspect of our work: our main focus in this paper is on verifying, using an SMT solver~\cite{de2008z3},
that the combination of several middleboxes enforces (\ie implements) a given invariant.

\subsection{Desired Network Properties}
\label{subsec:modeling:invariants}
We focus on three classes of invariants that address some of the core correctness issues plaguing networks:

\begin{compactdesc}
\item \textbf{Packet-level reachability and isolation between endhosts.}
This is the most straightforward network invariant: can two hosts exchange packets? In most cases we want to ensure that two hosts can exchange packets, but there are scenarios where isolation is crucial and here we want to ensure that the hosts cannot exchange packets. 
\item \textbf{Packet-level reachability and isolation between endhosts, with learning.}
This is a variation of the above invariant, where there is an asymmetry in that, for example, we might want to allow host $a$ to initiate contact with host $b$, but not allow host $b$ to initiate contact with host $a$. But once contact is properly initiated, we want two-way reachability.
\item \textbf{Content-level reachability and isolation between endhosts.}
One of the most interesting consequences of middleboxes is that the content of a host can be
leaked to another host (as through a cache) even when these hosts cannot exchange packets.
Therefore we also consider prevention of content exchange between two hosts (and this condition need not be symmetric; content from host $a$ might be allowed to reach host $b$, but not vice versa).
\end{compactdesc}

This is a very restricted class of invariants, but they can be used to address slightly more general questions, such as Traversal ({\it do packets going between source A and destination B always go through a particular element or link?}) and Preconditions ({\it are packet bodies modified before being processed by a particular middlebox?}). However, our current approach is not able to address invariants that address issues such as quantity ({\it how many packets can be sent between hosts?}), performance ({\it  do packets travel over uncongested links?}), or content ({\it are packets containing a certain string delivered?}), since these would require detailed consideration of each packet in the network, not just understanding broad classes of network behavior. Further, our choice of
invariants imply conditions on particular source-destination pairs (requiring for instance that source $a$ not communicate with source $b$) rather than applying to more general network-wide properties (\eg all source-destination pairs in a network use disjoint
paths).

\subsection{Middlebox Behavior}
\label{subsec:modeling-mboxes}
Because we are concerned with a limited class of \eat{behavioral }invariants, we need not consider fully detailed models of middleboxes. In fact, our invariants can be checked using relatively simple models that summarize what {\it possible} set of behaviors the middlebox might take for packets with a given header.  We make this more precise below.

We start with a few basic definitions: $P$ is the space of packets, $P^*$ is the space of all packet sequences, $H$ is the space of packet headers, and $MB$ is the set of middleboxes (including learning switches).
In this paper we assume middleboxes have a single output port.\footnote{A multiport middlebox can be modeled as a single-port middlebox followed by a multiport router.} Further middleboxes can depend on the entire packet (including the payload) and
on the history of packet arrivals\footnote{Middlebox state is derived from this history.}. A middlebox $m$ can be more formally represented as:
%
            \begin{align*}
                m: P\times P^* \rightarrow A \subseteq P
            \end{align*}
 where $A$ represents the middlebox's action on the packet: given a packet and a packet history, a middlebox can produce zero or more packets (which is why the range of the $m$ is not a single packet but a set of packets).

However, given our limited set of invariants, as we show later (through the success of our approach), we can make do with a {\it reduced model} that does not require detailed knowledge of the middleboxes decision process. This reduced model considers only how the behavior depends on the headers using a function $rm$:
            \begin{align*}
                rm: H\times H^* \rightarrow A \subseteq H
            \end{align*}
The reduced model does not prevent us from considering middleboxes whose behavior depends on the packet body, it merely takes the union of all such body-dependent behaviors and does not try to model which packet bodies elicit which behavior. This means that in order to model a middlebox we need not understand the details of its implementation, but only the broad outlines of the kinds of behaviors it supports.\footnote{However, our formulation includes both the general behavior of a middlebox and its current configuration; that is, firewalls have generic behavior, but also specific ACLs that determine which packets they drop. For simplicity we do not distinguish between the two here.}

Note that these reduced models of middleboxes are often quite simple.
Firewalls either forward a packet unchanged (if allowed), or block (if disallowed by an ACL), or forward conditionally if a hole has been punched by a packet in the opposite direction. Similarly, a cache either forwards a request or returns a response depending on whether it has a previously cached copy of the requested content.  Thus, we assume that these reduced forms can be specified in a limited grammar (described in \S\ref{sec:system:modeling}) and are equivalent to finite state machines.

Even with these reduced models, verification requires analyzing the entire network. However, as we argue later, this can be avoided due to the fact that many existing middleboxes (including firewalls) only depend on state pertaining to a particular
flow. We would like to focus on middleboxes where  $rm(h, S|_{flow(h)}) = rm(h, S)$ where $rm$ is a reduced middlebox function, $h \in H$, $S \in H^*$ and $S|_{flow(h)}\subseteq S$ is the sequence of headers which belong to the same flow as $h$ (the definition of flow can be arbitrary, as long as membership in a flow is a deterministic function of the header). This would effectively allow us to treat
a given middlebox as several parallel middleboxes, one per flow. However, it turns out that this simple definition is overly restrictive, and we need a more general definition as follows.
We  say a middlebox is {\em  Flow-Parallel} (FP) if and only if:
\begin{align*}
    \forall h,\ S \exists S' \supseteq S|_{flow(h)} s.t. rm(h, S'|_{flow(h)}) = rm(h, S)
\end{align*}
What this awkward definition means is that for every packet history, there is a possible flow history that can reproduce the same behavior. In short, the middlebox can never behave in a way that is inconsistent with a possible history of just that flow; all possible behaviors on a flow can be exhibited just by looking at the single flow. The pertinent example here is that an FP cache never returns content that was in the cache due to some other flow's previous request if it wouldn't have returned content if it had been requested by that flow.

\subsection{Pipeline Invariants}
\label{subsec:pipeline}

Along with the isolation invariants described in \subsecref{modeling:invariants}, we also need to check pipeline invariants. A pipeline invariant takes the form: {\it all incoming packets with headers belong to some $I\subseteq H$ must have passed through the sequence of middleboxes $mb_1, mb_2, mb_3,...$ before being delivered by the network}. Note that these invariants could refer to physical instances of middleboxes (e.g., packets must traverse this particular middlebox) or
a class of middleboxes (e.g., packets must traverse a firewall).  We assume that all packet headers belonging to the same flow are processed by the same pipeline (which can be enforced, as discussed below).

As we discuss in Section \ref{sec:system}, we can check these invariants using slight extensions to current verification tools.
This is possible by breaking the network into the static-datapath components and the dynamic-datapath components. A packet entering a static-datapath portion of the network (either from a middlebox,
or from an ingress port) emerges at an output port in $O$ or at a middlebox in $MB$ with perhaps a modified header. This behavior is described by a ``transfer function" $T$ which can be efficiently computed
by current verification tools, and when iterated will produce the pipeline that results from a given input packet.  As we discuss later, this is sufficient to scalably verify the pipeline invariants in large networks.

\subsection{Stronger Enforcement Conditions}
\label{subsec:modeling:stronger}
An operator's goal is to design their network --- including the network topology, where middleboxes are placed in the network, and how they are configured --- that can enforce their desired invariants. In this paper
we examine how we can efficiently verify that a particular network achieves this result. It is simple to efficiently verify pipeline invariants. However, these techniques do not apply
to our other invariants, whose enforcement depends in more detail on the behavior of middleboxes. For these invariants we rely on particular forms of pipelines to help scale verification.

Consider a simple pipeline $\Pi$: a sequence of middleboxes $\Pi = \{m_1, m_2, m_3, \ldots, m_k\}$ (for simplicity, we ignore the possibility that intervening routers rewrite any packets and assume they merely forward them).\footnote{More complicated pipelines can be DAGs, not just a single sequence; that is, the pipeline can branch at points depending on the actions intervening middleboxes and routers. Our tools deals with such cases, but for simplicity we ignore this possibility in this section.} This pipeline is similar to a middlebox: it maps an incoming packet and a history into one or more outgoing packets. However unlike a middlebox, the computation for a pipeline depends not
just on the history of packets that traverse the pipeline, but also on packets that are received and processed by any of the constituent middleboxes. More formally:
\begin{align*}
    \Pi: H\times S_n \rightarrow A \subseteq H
\end{align*}
where $S_n$ is the sequence of all packets sent in the network. Analyzing such pipelines is expensive since it requires considering the behavior and possible histories of all the constituents of a network.

Similar to flow-parallel middleboxes, we say that a pipeline is ``rest-of-network-oblivious'' (RONO) if and only if:
\begin{align*}
  \forall h,\ S_n \exists S' \supseteq S_n|_{flow(h)}\ s.t.\ \Pi(h, S'|_{flow(h)}) = \Pi(h, S_n)
\end{align*}
where $S'|_{flow(h)}$ is the history restricted to the flow defined by header $h$ as above, and whose packet flow through the entire pipeline. As noted previously, the isolation invariants we focus on are all stated in terms of pairs of endhosts (and can thus
be naturally extended to a set of flows). It is now simple to see that given an invariant $I$ involving hosts $a$ and $b$, and a set of RONO pipelines $\Pi_1\ldots \Pi_n$ connecting $a$ and $b$ (each applying to a different set of headers), $I$ holds for the entire network if and only if
$I$ holds for all pipelines $\Pi_1$ to $\Pi_n$.

Thus if one can enforce the desired invariants in a network using RONO pipelines, then one need only verify invariants on the pipeline in isolation. While RONO and flow-parallelism are closely related, somewhat surprisingly not all compositions of FP middleboxes are RONO. In \S\ref{sec:tractability:rono} we derive conditions under which compositions of FP middleboxes are guaranteed to be RONO. Isolation invariants can thus be verified quickly by
checking that (a) all middleboxes in the pipelines connecting the endhosts are flow-parallel, (b) the pipelines themselves are RONO and (c) the invariant holds on each pipeline. The first two steps are relatively
simple static checks that depend only on the middlebox model specification and the last verification step generally scales with the length of the pipeline and policy size (\ie the number of invariants) rather than the size of the network. We evaluate scalability empirically in \S\ref{sec:eval}.

\section{System Design}
\label{sec:system}
Based on the previous theory, we have implemented a system to verify invariants of the type described in \S\ref{subsec:modeling:invariants} in a given network.
Our system uses Z3~\cite{de2008z3}, a state-of-the-art SMT solver, as an oracle that can prove theorems of the appropriate form.
For verification we require two inputs: (a) middlebox models written in a restricted language based on Python (\S\ref{sec:system:modeling}); 
(b) network topology information including routing tables, middlebox configurations and end host metadata. The system converts these inputs to a suitable form, adds additional assertions (\S\ref{sec:system:netassert}) describing the physical behavior of the network, as well as the invariant being checked, and produces the input supplied to Z3. Given this input, Z3 returns \texttt{unsat} (indicating that the input assertions can never be satisfied), \texttt{sat} (indicating that a satisfiable assignment
was found) or undefined (indicating that the check timed out). The system interprets this return value to determine if the invariant holds. We describe each of these steps below.

\subsection{Modeling Middleboxes}
\label{sec:system:modeling}
\definecolor{codegray}{rgb}{0.5,0.5,0.5}
\lstdefinestyle{mypy}{
  float=tb,basicstyle=\footnotesize,
    stringstyle=\ttffamily,numberstyle=\tiny\color{codegray},breaklines=true,numbersep=5pt,                  
    showspaces=false,                
    showstringspaces=false,
    showtabs=false,                  
    tabsize=2,language=Python,
    numbers=left,
    linewidth=0.9\columnwidth,
    breakatwhitespace=true,
    belowskip=0pt,
    aboveskip=0pt,
    abovecaptionskip=2pt,
    belowcaptionskip=0pt,
    frame=lines,
    columns=fixed,
    captionpos=b,
    basewidth=0.6em}
    
\begin{lstlisting}[style=mypy,
                   caption=Model for a learning firewall, 
                   label=code:lfw]
def LearningFw(node, acl, flows):
  p = recv(node)
  if acl[p.src(), p.dest()]:
    send(node, p)
    flows.set((p.src(), p.dest(), p.src_port(), p.dest_port()), True)
  elif flows[(p.dest(), p.src(), p.dest_port(), p.src_port())]:
    send(node, p)

acl = ConfigMap((Address, Address), Bool)
flows = Map((Address, Address, Int, Int), Bool)
LearningFw(f, acl, flows)
\end{lstlisting}

SMT solvers cannot always prove (or disprove) fully general theorems\footnote{First-order logic is undecidable in general and we must restrict ourselves to
formulas in a decidable fragment. See \S\ref{sec:tractability:decidability} for details.} and we must limit the complexity of our input to Z3.
We therefore require that middlebox models used by our system are restricted so that:
\begin{asparaenum}
\item Models are expressed so they are loop-free and all received packets are processed in a fixed number of steps.
\item Models only access local state which is naturally true for existing middleboxes, which are physically distinct and must use the network to share state.
\item Models are expressed using a limited set of actions: they can receive packets, check conditions, send packets and update state. 
\item Models must be deterministic for a given packet and history. For flow-parallel (\emph{FP}) middleboxes we require that 
the modeled behavior be identical for a given packet and all histories with the same flow-restricted history, \ie $m(h, S) = m(h, S')$ for all $S, S' \in H^*$ 
with $S|_{flow(h)} = S'|_{flow(h)}$ and FP middlebox $m$. While existing implementations can be non-deterministic (\eg NATs that assign ports in order of flow initiation), these have
equivalent, semantically correct, deterministic versions (for instance a NAT that uses flow hashing to assign ports) for which our invariants hold if and only
if they also hold in the non-deterministic case.
\end{asparaenum}

Users specify middlebox models (which are general and can be reused for different networks) using a subset of Python that allows users to:
\begin{asparaitem}
\item Read and set values from instances of \texttt{Map} objects, which behave like dictionaries or hash maps.
\item Read values from \texttt{ConfigMap} objects, which hold configuration information.
\item Call uninterpreted functions with finite codomains (\ie returns one of a finite set of values).\footnote{In our current implementation we do not check that uninterpreted
functions have finite codomains.}
\item Use conditionals \texttt{if}, \texttt{else}, \texttt{elif}.
\item Construct a \texttt{packet} and set packet field values.
\item Call the \texttt{recv} function to receive a packet.
\item Call the \texttt{send} function to send a packet.
\end{asparaitem}

As an example, consider the model for a {\em learning firewall} shown in Listing~\ref{code:lfw}.  The firewall can forward received packets either because
this is explicitly allowed by the firewall policy (line 3, where we check to see if the packet is allowed by the list of ACLs) or because a previously allowed packet
established flow state (in line 5, we modify the \texttt{Map} flows to record what packets have been seen, which we then check in line 6). The model itself is specified
by the function definition (lines 1--7). Lines 9--11 show how this model can be initialized for a node $f$. 

\begin{lstlisting}[style=mypy,
                   caption=Model for a DPI middlebox, 
                   label=code:dpi]
dpi = Function([Body], Bool)                   
def DPIFw(node):
  p = recv(node)
  if dpi(p.body()):
    send(node, p)
\end{lstlisting}

Listing~\ref{code:dpi} shows an example where an uninterpreted function \texttt{dpi} (defined on line 1) is used. \texttt{dpi} accepts a packet body and returns a boolean 
and hence has a finite codomain of size $2$ (\ie it returns true or false).

The system translates these models into an equivalent set of formulas in temporal logic that we supply to Z3. Figure~\ref{fig:fwformula} shows the formulas for an instance ($f$)
of the learning firewall.

The translation works by performing a depth first traversal of the abstract syntax tree (AST) to find all calls to \texttt{send} or \texttt{set} (henceforth referred to as
``actions'') and the path leading to these calls. The path is converted to an appropriate path constraint and we output assertions of the form $\text{\it action} \implies 
\text{\it path constraint}$, essentially requiring that if an action (\eg a packet is forwarded) is executed then all conditions leading up to it must hold.  

Our model description could produce several equivalent sets of formulas. Later in \S\ref{sec:tractability} we use one such equivalent formulas to prove that our formulation
is decidable. We chose this particular form of formulas based on how long it took Z3 to produce a proof for these formulas.

\begin{figure}[tb]
\centering
\begin{align*}
    send(f, n, p, t) &\Rightarrow \exists t', n' .\, recv(n', f, p, t') \land t' < t \\
                                                  &\ \land (acl(s(p), d(p)) \\
                    &\lor (acl(s(p),d(p)) \\
                                          &\  \land flows(d(p),s(p),dp(p),sp(p), t')))\\
    flows(s, d, s_p, d_p, t) &\Rightarrow\\
    &(\exists t', n, p.\, recv(n, f, p, t') \land t' < t\\
                            & \land acl(s(p), d(p)) \land s = s(p)\\
                            & \land d = d(p) \land d_p = dp(p) \\
                            & \land s_p = sp(p))
\end{align*}
\vspace{-0.21in}
\caption{Formulas generated from Listing~\ref{code:lfw}. The left side of the first two formulas have a universal
quantifier, \ie each is preceded by $\forall n, p, t$ etc. $f$ is a symbolic node representing the firewall.}
\label{fig:fwformula}
\vspace{-0.1in}
\end{figure}
\subsection{Network Transfer Functions}

\label{sec:system:transfer}
We also place a few restrictions on the topology and forwarding state for networks we verify, in particular we require that the networks under consideration be:
\begin{asparaenum}
\item Forwarding loop free, this is required to ensure that the formulas supplied to the SMT solver are decidable.
\item Have no black holes (\ie routers and switch forwarding tables be setup such that packets are always forwarded to their destination). 
\end{asparaenum}

We leverage VeriFlow~\cite{khurshid2012veriflow} to both check that the input topology and forwarding tables meet the previous requirements and to produce a forwarding graph that we
can then convert to a set of ``composition assertions'' that we add to the theorem provided to Z3. Along with the topology and forwarding table we also accept configuration for middleboxes
and endhosts. This configuration specifies the type of the node (which we use to create a new instance from the appropriate model) and any configuration that the model might depend on. For end hosts
this configuration also specifies the set of addresses assigned to a specific host (we assume that hosts are honest, \ie they do not send packet with spoofed addresses, this can be easily enforced 
at the first hop).

Once these models are instantiated we query the forwarding graph to determine the possible pipeline(s) traversed by a packet sent from one end host to 
another.  We translate these pipelines into composition constraints of the form 
$send(a, n, p, t) \land (d(p) = ip_b) \land (s(p) = ip_a) \implies (n = f)$. The previous composition constraint indicates that packets leaving host $a$ with destination address
$ip_b$ and source address $ip_a$ are sent to the firewall next. We refer to this collection of instantiated middlebox models and composition constraints as the \emph{network model}

\subsection{Other Assertions}
\label{sec:system:netassert}
Next we add to the network model some basic axioms describing the universe in which the network operates. These axioms (Figure~\ref{fig:general}) state that:
\begin{asparaenum}
\item The network has no local loops (\ie no packets with the same source and destination address).
\item Any packet received at a node $n'$ from node $n$ at time $t$ was sent by $n$ at an earlier time $t'$.
\item Time is represented by positive numbers.
\end{asparaenum}

\subsection{Verification}
\label{sec:system:verification}
Finally, we add to the network model one or more variables (representing packets) and assertions on these variables (encoding conditions that should or should not hold if the
invariant holds) to generate an input for Z3. Given this input Z3 either returns a valid assignment for the variables such that the supplied assertions hold under the network model 
(\texttt{sat}) or that no such assignment exists (\texttt{unsat}). The variables and assertions added for each invariant are:
\begin{asparaitem}
\item \textbf{Node Isolation}: To check node isolation between nodes $a$ and $b$, we add a variable representing a packet ($p$) and assertions requiring that the packet was sent by $a$ and
later received by $b$. If the solver returns \texttt{unsat} no such packet can exist and the nodes are isolated. Note, that node reachability is the negation of this and is true whenever the
solver returns \texttt{sat}.
\item \textbf{Flow Isolation}: Flow isolation is verified by adding an additional assertion to Node Isolation: the additional assertion states that $b$ has never before sent a packet to $a$.
\item \textbf{Data Isolation}: We rely on a pseudo-field on our packet to indicate what machine data originated on. Models for caching firewalls are expected to preserve this field. Given this
pseudo-field we can check that $a$ never accesses data from $b$ by proving that there does not exist a packet ($p$) such that $p$ is received at $a$ and has origin $b$.
\item \textbf{Node Traversal}: The node traversal invariant requires that all traffic from host $a$ to host $b$ pass through some middlebox $m$. This can be proved by showing that if
$m$ neither nor receives any packets then $a$ and $b$ are (node) isolated from each other.
\end{asparaitem}
\begin{figure}[tb]
\centering
\begin{flalign*}
   send(n, n', p, t) &\implies n\neq n'\\
   send(n, n', p, t) &\implies s(p) \neq d(p)\\
   recv(n, n', p, t) &\implies \exists t'.\, send(n, n', p, t') \land t' < t\\
   send(n, n', p, t) &\implies t > 0\\
   recv(n, n', p, t) &\implies t > 0
\end{flalign*}
\caption{Basic network axioms.}
\vspace{-0.2in}
\label{fig:general}
\end{figure}

We can also use Node Isolation with an additional constraint to measure the number of host pairs which can potentially use a link (path in our case), we call this \emph{link traversal}. To measure
link traversal we run $2\times {n \choose 2}$ node isolation checks with the added constraint that the packet must have gone over the link. We then return the number of cases in which Z3 returns \texttt{sat}
for this check.

\subsection{Enforcement}
\label{sec:system:enforcement}
Our models are not automatically derived from implementations and hence it is possible (due to bugs) that the implementation for a middlebox deviates from its model. We 
address this by providing a runtime mechanism for detecting instances where the implementation's behavior deviates from what is allowed by the model, we call this
process \emph{enforcement}.

Since our models are specified as simple Python programs (\S\ref{sec:system:modeling}) they can be executed as long as we generate an implementation for uninterpreted functions.
Since, uninterpreted function have finite codomains, we provide a simple implementation where we execute the model once for each value in an uninterpreted function's codomain. Given
this implementation our enforcement strategy is simple: when a packet is received at a middlebox, we send a copy of the packet to the enforcement code, which generates all possible
outputs that are allowed by the model. We then compare the middlebox's output to these possibilities and report a deviation when no match is found.
\eat{
Further, while our enforcement strategy is an over-approximation and will not detect incorrect (but consistent behavior), it is designed to be used in addition to more expensive static checks
or automatic test packet generation~\cite{zeng2012automatic} techniques. We envision one could develop static verification techniques to check that an implementation is fateful to a
supplied model, however developing such techniques is left to future work.
}
\eat{\begin{outline}
\1 Current implementation built on top of Z3, however most of the theory is agnostic to proof technique used.
\1 Modelling Middleboxes
    \2 We use a small DSL with limited grammar to express middlebox code.
        \3 Why?
        \3 Grammar
    \2 Translated to first-order logic with some additional theorems: quantifiers, uninterpreted functions, etc.
        \3 Show an example or two.

\1 Network transfer functions
    \2 Derived from the forwarding tree built by VeriFlow.
        \3 Loosely coupled.
    \2 Translating the forwarding tree.
    
\1 Putting it all together for verification

\1 Enforcement
    \2 Models are written independently of middlebox applications.
    \2 A common question: how can we be sure the models match the applications. 
    \2 Might be possible to do this statically, left to future work. We however have a dynamic version.
    \2 Middlebox DSL can be expanded to executable code (instead of first-order logic).
        \3 Code runs for each packet, used to check if given the current packet (and all previous packets) will model output a packet or not.
\end{outline}}
\section{Theoretical Analysis and Decidability}
\label{sec:tractability}
Next we try and answer two questions about our formulation: are our formulas decidable (\S\ref{sec:tractability:decidability}) \ie can Z3 solver find a satisfiable assignment (or prove that the formulas are unsatisfiable) in a finite number of steps, and secondly conditions under which combination of middleboxes will result in RONO pipelines (\S\ref{sec:tractability:rono}).



\subsection{Decidability}
\label{sec:tractability:decidability}

\begin{figure}[tb]
\centering
\begin{flalign*}
    snd(e) \land s(e) = f &\implies t(cause(e)) < t(e)\\
                    &\land rcv(cause(e)) \\ 
                    &\land d(cause(e)) = f \\
                    & \land p(cause(e)) = p(e) \\
                    & \land acl(s(e1), d(e1)) \\
    rcv(e) \lor s(e) \neq f &\implies cause(e) = e 
\end{flalign*}
\caption{EPR-F formulas equivalent to \\
$send(f, n, p, t) \Rightarrow (\exists t', n'. recv(n', f, p, t') \land t' < t \land acl(s(p), d(p))$.}
\label{fig:aclfwevformula}
\vspace{-0.1in}
\end{figure}

In general, first order logic is undecidable. However we show that when we restrict our models (\S\ref{sec:system:modeling}) and topology (\S\ref{sec:system:transfer}) as described
previously, we get formulas that lie in a decidable fragment of first-order logic.
This fragment is a simple extension of ``effectively propositional logic" (\emph{EPR}). EPR is one of the fundamental decidable fragments of first-order logic~\cite{JAR:PiskacMB10}.
An EPR formula is a set of function symbol free $\exists^*\forall^*$ premises (assertions) and a function symbol free $\forall^*\exists^*$ formula as a consequence (whose negation 
is added as an assertion before checking for satisfiability). Z3 and other SMT solvers use algorithms that are guaranteed to terminate for this fragment. 

In our case, the formulas obtained from modeling middleboxes (\eg Figure~\ref{fig:fwformula}) as well as one of the network axioms (assertion 3 in Figure~\ref{fig:general}) are $\forall^*\exists^*$ premises and hence not in EPR. Such
premises may result in the formula being undecidable (which would cause the SMT solver to timeout). Therefore our case requires a more expressive logic that we call EPR-F. EPR-F extends EPR to allow restricted
unary functions. To ensure that the formulas are decidable, EPR-F requires that unary functions have certain closure properties such that some finite composition (including composition of the function with itself)
must result in an idempotent function, \eg for a EPR-F formula with a single function $f$ there must exist $k$ such that $f^k(x) = f^{k-1}(x)$.
Note that ``non-cyclic" function symbols that go from one type to another\footnote{Assuming no function go back.} can be employed freely~\cite{Korovin}.
A fragment similar to EPR-F was employed in \cite{Itzhaky:2014:MRH:2535838.2535854}, which also shows that EPR-F formulas can be reduced to pure EPR. 

One can translate our models (\eg Figure~\ref{fig:fwformula}) to EPR-F using the following steps and then use
the fact that our topology and middleboxes are loop-free to prove that all introduced functions are idempotent:
\begin{asparaenum}
\item Reformulate our assertions with ``event variables'' and functions that assign properties like time, source and destination to an event. We use predicate function to mark events
     as either being sends or receives.  \eat{For example we translate the formula $\forall n, n', p, t\, . recv(n, n', p, t)\implies \exists t'\,. send(n, n', p, t') \land t' < t$ to the equivalent formula
     $\forall e\, . rcv(e) \implies \exists 'e\, . snd(e) \land s(e) = s(e') \land d(e) = d(e') \land t(e') < t(e)$.}
\item Replace $\forall\exists$ formulas with equivalent formulas that contain Skolem functions instead of symbols. \eat{For instance the above formula is converted to 
$\forall e\, . rcv(e) \implies snd(cause(e)) \land \ldots \land t(cause(e) < t(e)$ where $cause$ is an uninterpreted unary function. Further, we add a second constraint saying that $\forall e\, . snc(e) \implies cause(e) = e$ (or 
equivalently that $cause(e)$ is idempotent for events that are not $rcv$ events. }
\end{asparaenum}
For example the formula $\forall n, n', p, t\, . recv(n, n', p, t)\implies \exists t'\,. send(n, n', p, t') \land t' < t$ is translated to the equivalent formula
$\forall e\, . rcv(e) \implies snd(cause(e)) \land \ldots \land t(cause(e)) < t(e)$ and we add a second assertion $\forall e\, . snd(e) \implies cause(e) = e$ to ensure that $cause(e)$ is idempotent for send events.
Figure~\ref{fig:aclfwevformula} shows another example of this reformulation for a simple ACL firewall (with no learning action).

Finally, we need to show that these newly introduced formulas have the desired closure properties. Note first that each cause functions is idempotent: they are the identity function for either send
or receive events and when not the identity function $cause$ links a send to a receive.
Furthermore, when applied to an event $e$ the value of this function (when not idempotent) is always another event such that $d(e) = s(e)$. Since we assume that we have loop free forwarding this must terminate
in a finite number of steps (when we would have reached the edge of the network) and therefore the newly introduced functions meets the requirement for being in EPR-F.

\begin{table}[tb]
\centering
\begin{tabular}{|l|r|r|}
\hline
Hosts & Firewall & Content Cache\\
\hline
$500$ & $43.87$s & $282.94$s \\
$1000$ & $384.15$s & $4264.16$s\\
$1500$ & $1736.70$s & $19819.72$s\\
\hline
\end{tabular}
\caption{Time to verify invariants in larger
networks.}
\vspace{-0.1in}
\label{tab:largeverification}
\end{table}
\subsection{RONO Pipelines}
\label{sec:tractability:rono}
\begin{figure*}[tb]
\subfigure[]{
  \centering
  \includegraphics[width=0.31\textwidth]{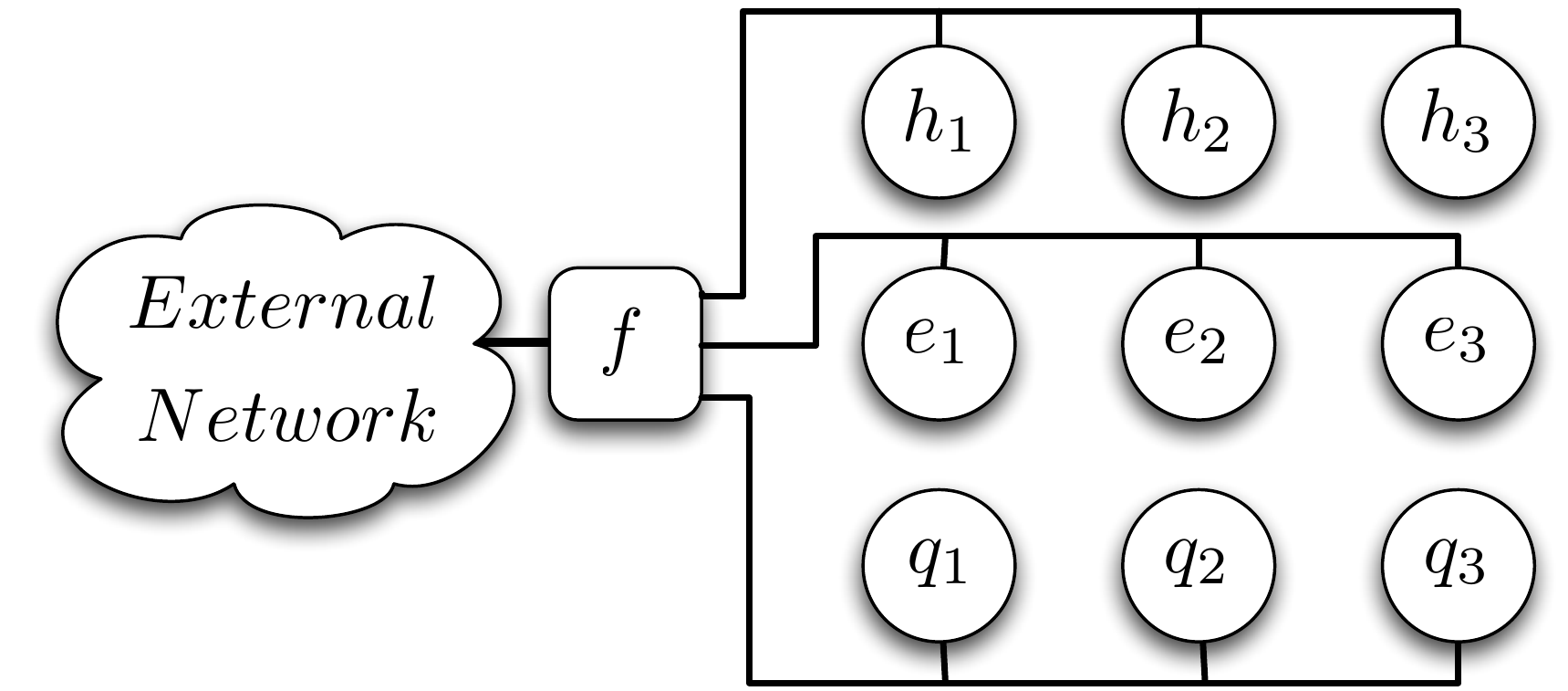}
   \label{fig:enterprise}
}
\hfill
\subfigure[\label{fig:singlefw_pipe}]{
  \centering
  \includegraphics[width=0.31\textwidth]{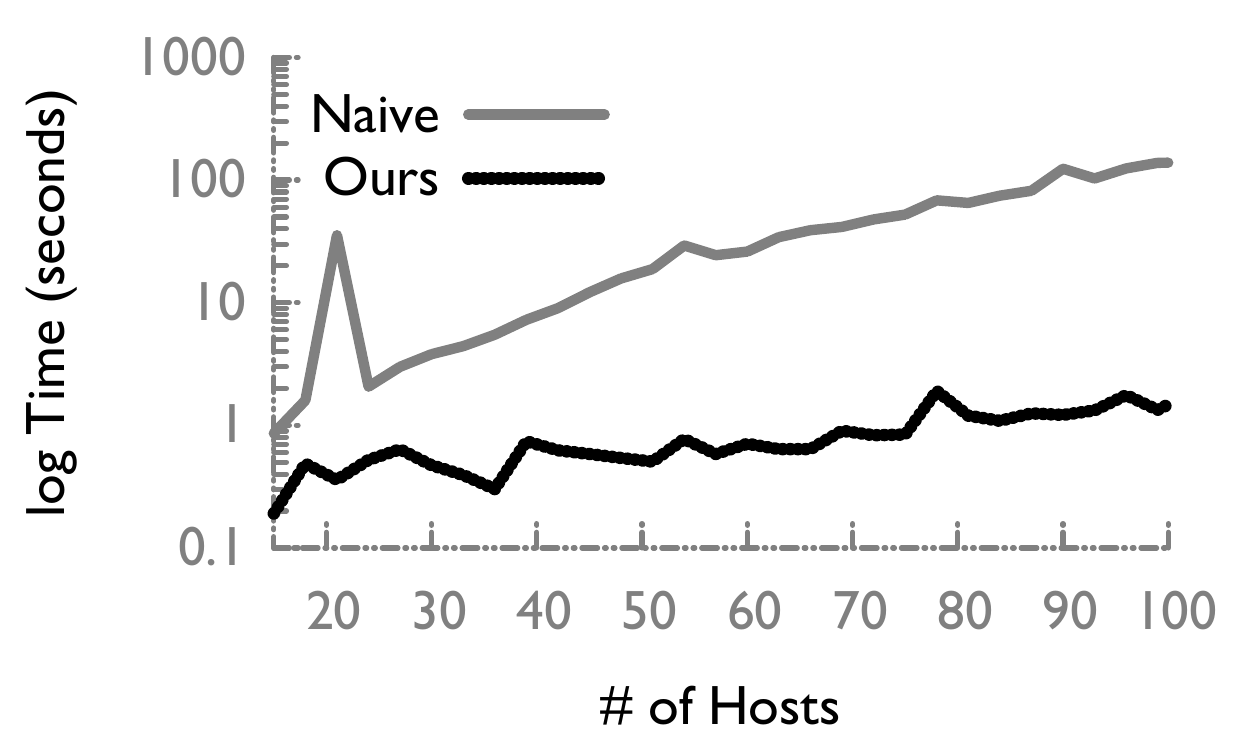}
}
\hfill
\subfigure[]{
  \centering
    \includegraphics[width=0.31\textwidth]{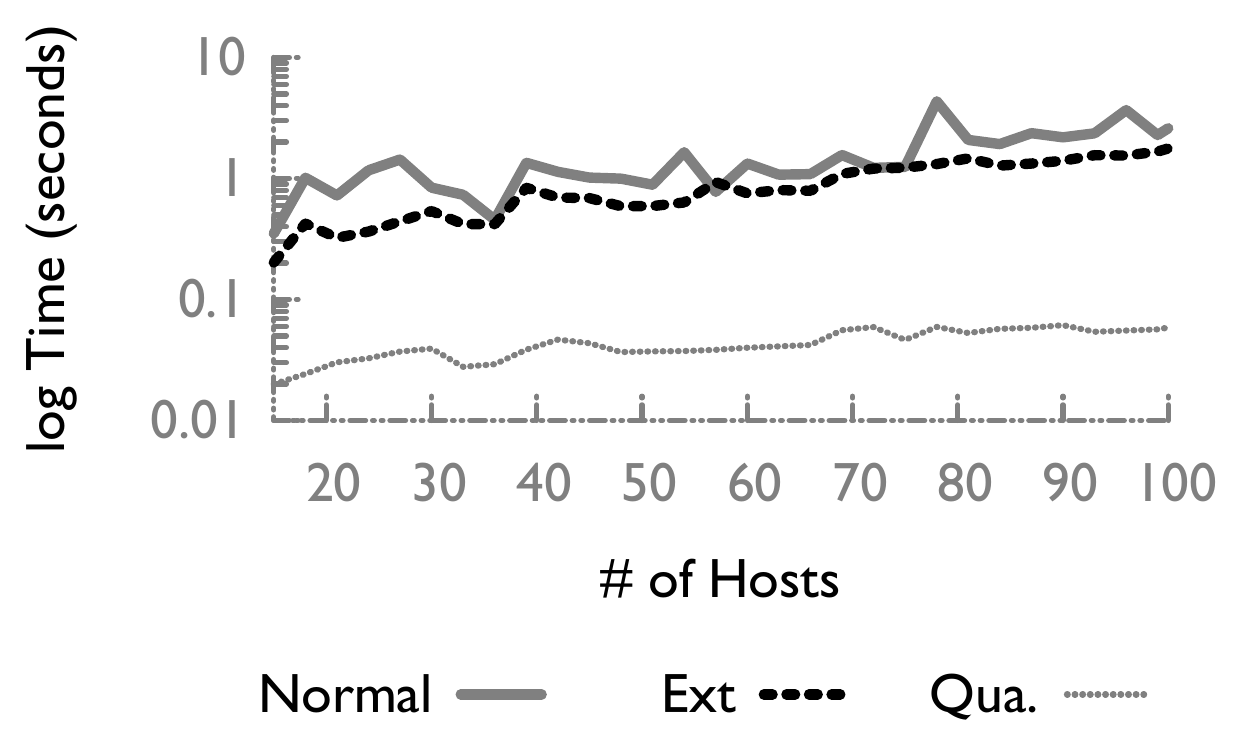}
   \label{fig:singlefw_pipe_break}
}
\caption{\subref{fig:enterprise} Topology for an enterprise network: $f$ is a whole punching firewall, $e_k$, $h_k$ and $q_k$ are different classes of hosts; \subref{fig:singlefw_pipe} 
average verification time per pipeline; \subref{fig:singlefw_pipe_break} breakdown of pipeline verification time per invariant type.}
\vspace{-0.2in}
\label{fig:singlefw_all}
\end{figure*}
In this section we look at conditions under which network pipelines are RONO. To start with define composition of two middleboxes (or pipelines, which have the same form)
$\left\{m_1, m_2\right\}$ to be the single middlebox representing the case where all outputs from $m_1$ are sent to $m_2$. More formally we define the composition function $C$ as:
\begin{align*}
C(m_1, m_2)(p, S) &= \left\{m_2(p', S) \, \vert \, p'\in m_1(p, S)\right\}
\end{align*}
where $S$ is the combination of both $m_1$ and $m_2$'s histories.

As described previously, we say that a pipeline $\Pi$ is RONO if and only if
\begin{align*}
  \forall S_n \exists S' \supseteq S_n|_{flow(h)}\ s.t.\ \Pi(h, S'|_{flow(h)}) = \Pi(h, S_n)
\end{align*}
\ie a single middlebox equivalent to the composition of all the middleboxes (and routers and switches) in a pipeline is flow parallel. 
In this section we analyze cases when pipelines are RONO and cases where pipelines are not. Note, our analysis here is conservative, \ie we 
find sufficient conditions for pipelines to be RONO and it is possible that other pipelines are also RONO.

For the statements below we focus on the composition of flow parallel middleboxes. While one can in theory construct RONO pipelines out of middleboxes that are not flow parallel,
the precise conditions for doing so depend on the behavior of the middleboxes in question and are hard to generalize. 

\textbf{All pipelines containing a single flow-parallel middlebox are RONO}: this follows trivially from the definition of flow-parallel middleboxes and RONO pipelines.

Based on the previous result we state all subsequent results in terms of RONO pipelines.

\textbf{Not all compositions of RONO pipelines are RONO}: \eat{\noteori{it seems that the example is for something different than the title. The example below is of a mb added to a pipeline and not of two pipelines. Maybe we should explicitly say that there is no difference? also $C$ is only defined two mb's. Did I miss anything?}} We show this by presenting a simple counter-example. Consider the middlebox $m(p, S) = \left\{ c \right\}$
where $c\in P$ is a packet with source $ip_{m}$, destination $ip_{m}'$, source port $s_m$ and destination port $d_m$, \ie $m$ outputs a constant packet $c$ for any input.
Since the definition of $m$ is independent of $S$ we have $\forall S\in H^{*}.\ m(h, S|_{flow(h)}) = m(p, S)$ and $m$ is trivially flow parallel. Now consider the composition of
$m$ with a RONO pipeline $\Pi$. Since $m$ outputs the same packet $c$ regardless of the received packet, for $\Pi$ flows are indistinguishable and hence it cannot partition the history of 
received packet. $C(m, \Pi)$ is therefore not RONO despite both $m$ and $\Pi$ being RONO.

We say a RONO pipeline $\Pi$ is flow preserving if and only if:
\begin{align*}
\forall S, h, h': flow(\Pi(h, S)) &= flow(\Pi(h', S)) \iff\\ 
&flow(h) = flow(h')
\end{align*}
that is $\Pi$'s action maps headers for packets belonging to the same flow to packets in the same flow (\ie a flow-preserving pipeline $\Pi$ is injective with respect to flows).

\textbf{The composition of two flow preserving pipelines is also flow preserving} -- the composition of two injective functions is injective.

\textbf{The composition of flow-preserving RONO pipelines is RONO}: This is obvious to see: the previous problem is a result of flows being merged, this is impossible given flow-preserving middleboxes. 

We thus see that any pipeline of flow-preserving, flow-parallel middleboxes is RONO.

\section{Evaluation}
\label{sec:eval}

We now evaluate our system's performance and demonstrate gains when compared to an alternative that model-checks the entire network (dubbed ``na\"ive approach''). 
First, we show that our system can be used to verify existing networks in reasonable time. 
We evaluate our system's performance and scalability both on enterprise and departmental networks that use stateful firewalls (\S\ref{sec:eval:firewall}) and on provider networks that use content caches (\S\ref{sec:eval:cache}). 
Next, we evaluate our system's performance on more general functionality and show that it can be used to check invariants in the presence of other kinds of middleboxes (\S\ref{sec:eval:genmb}).
Next, we evaluate the benefits of RONO (\S\ref{sec:eval:nonrono}). 
We close by evaluating our system's performance when verifying node-traversal invariants (\S\ref{sec:eval:extra}).

\begin{figure*}[tb]
    \centering
    \begin{minipage}{0.31\textwidth}
        \centering
        \subfigure[]{
             \includegraphics[width=\textwidth]{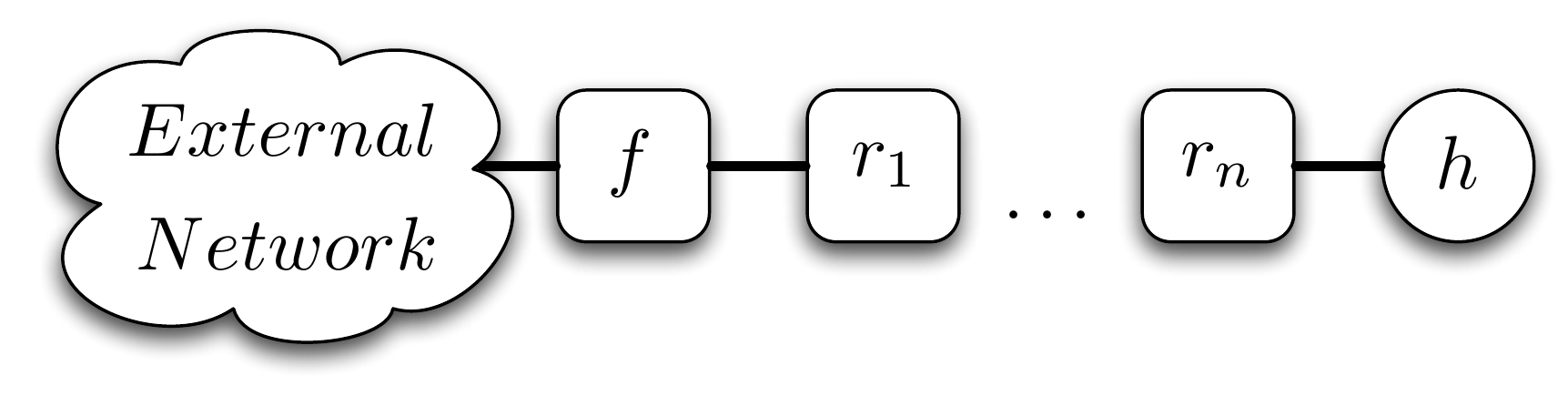}
             \label{fig:pathlength}
        }
        \subfigure[]{
            \includegraphics[width=\textwidth]{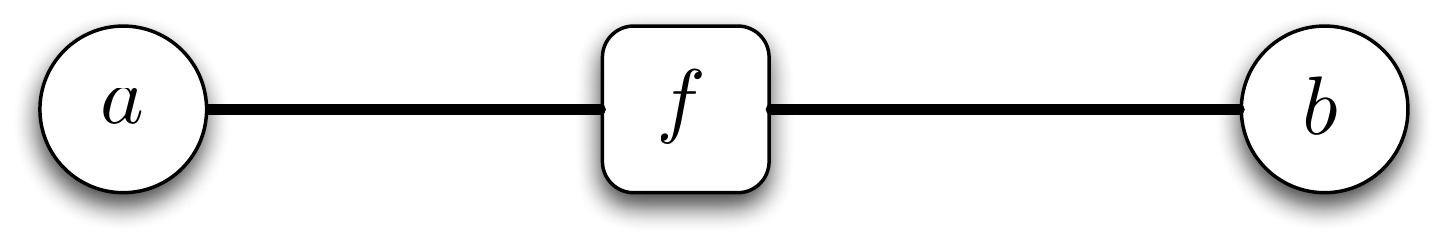}
            \label{fig:policyscaling}
        }
        \caption{\subref{fig:pathlength} Topology for pipeline-length scaling; \subref{fig:policyscaling} topology for policy-size scaling}
         \vspace{-0.2in}
    \end{minipage}
    \hfill
    \begin{minipage}{0.31\textwidth}
        \centering
        \includegraphics[width=\textwidth]{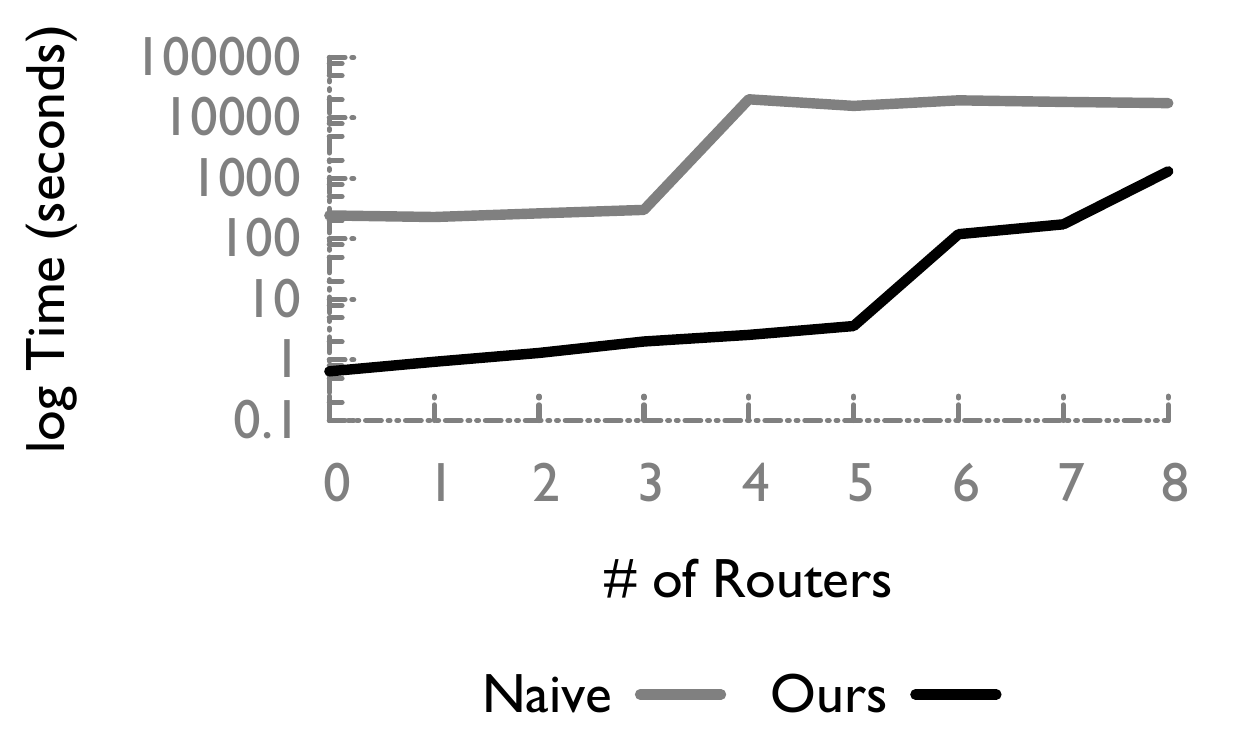}
        \caption{Average pipeline verification time with increasing pipeline length.}
        \label{fig:pathscaling}
         \vspace{-0.2in}
    \end{minipage}
    \hfill
    \begin{minipage}{0.31\textwidth}
        \centering
                \includegraphics[width=\textwidth]{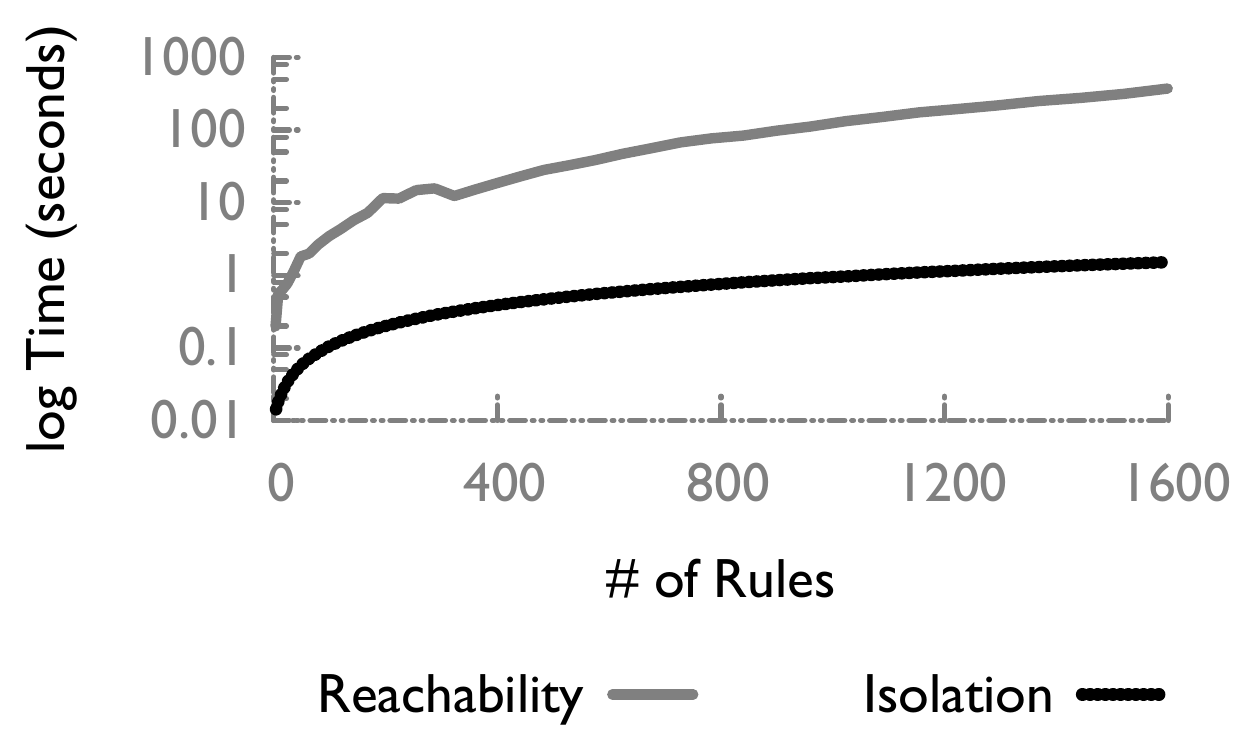}
         \caption{Pipeline verification time with increasing policy size.}
            \label{fig:policyscale}
         \vspace{-0.2in}
    \end{minipage}
\end{figure*}
\subsection{Stateful Firewalls}
\label{sec:eval:firewall}

{\bf Single FP Middlebox.}
We first consider the topology in Figure~\ref{fig:enterprise}, where one stateful firewall\footnote{They may be several redundant physical firewall devices for fault tolerance, but a single firewall
processes all traffic between the internal and external network.} protects a network that consists of three groups of hosts: \emph{Quarantined}, \emph{Normal},
and \emph{External-facing}. This arrangement is typical of small and medium-size enterprises and several departmental networks (including the network at UC Berkeley). 
We use our system to verify the following node-isolation and flow-isolation invariants:
\begin{asparaenum}
\item Quarantined hosts are node-isolated: No quarantined host can send or receive packets from the external network.
\item External-facing servers can both access and be accessed from the external network.
\item Normal hosts are flow-isolated: Any normal host is allowed to establish connections and communicate with nodes in the external network, but the external network cannot establish a connection with the host.
\end{asparaenum}

To implement these invariants, we configure the firewall with two rules denying access (in either direction) for each quarantined host, plus one rule denying inbound connections for each normal host. For our
evaluation, we consider a network with equal numbers of hosts of each group (\ie a third of the hosts are quarantined and a third are externally accessible). We configure our firewall correctly, and our results are for the case where all invariants hold. Note, however, that misconfiguring the firewall so that an invariant does not hold for a particular host merely places that host in a different group. For instance, suppose the firewall is misconfigured causing a particular quarantined host not to be isolated; from the point of view of verification complexity, this is similar to the situation where the host is a normal or external-facing host, and the firewall is correctly configured. Thus, our evaluation provides relevant timing information for both the case where an invariant holds and the cases where it is violated.

To start with, we measure the time taken to verify that the correct invariants hold for all the hosts in the network. Learning firewalls are flow-parallel, hence each pipeline that involves a host and the firewall is RONO and can be checked in isolation. Figure~\ref{fig:singlefw_pipe} shows pipeline verification time when we check each pipeline in isolation (``Ours'') and when we check the entire network (``Na\"ive''): for a moderately sized network with a $100$ hosts, our system takes about $1$sec per pipeline; hence, in the worst-case scenario where it runs on a single core, it checks all $100$ pipelines in about $100$ seconds, which is two orders of magnitude faster than the na\"ive approach. Table~\ref{tab:largeverification} shows pipeline verification time for our system, when we have larger numbers of hosts: for a network of $1500$ hosts, it is close to half an hour; the na\"ive approach does not terminate in useful time.

Even for our system, pipeline verification time increases with the number of hosts in the network, because the number of hosts affects the number of rules that are installed in the firewall, and these rules must be checked for each pipeline.
Figure~\ref{fig:singlefw_pipe_break} breaks down pipeline verification time per invariant, and we see that verifying node-isolation takes less time than verifying flow-isolation or reachability. This is because node-isolation is expressed as unsatisfiability, and Z3 has a known pathology where, for many problems, proving unsatisfiability is faster than finding a satisfiable assignment. While this is a known problem, the exact reason is not understood. Other solvers, for instance CVCLite, do not suffer from this limitation but have other idiosyncrasies.

\vspace{0.1cm}
{\bf Increasing Pipeline Length.}
In reality, hosts in Figure~\ref{fig:enterprise} connect to the firewall through a series of switches, routers, and perhaps other middleboxes that we have elided. Next, we look at two questions: (i) what is the cost of explicitly modeling switches/routers and (ii) how does verification scale with increasing pipeline length. We analyze both of these by changing the pipeline in Figure~\ref{fig:enterprise} by introducing a series of routers (or middleboxes that forward all received packets) as shown in Figure~\ref{fig:pathlength}.

Figure~\ref{fig:pathscaling} shows pipeline verification time for pipelines of increasing length. We find that it grows rapidly with increasing pipeline length, both for our system and the na\"ive approach. Even so, our system is two orders of magnitude faster. Fortunately, due to the additional latency added by each middlebox, we expect the number of middleboxes in a pipeline to be relatively small. Further, given that we require the forwarding configuration of switches and routers in the network under analysis to contain no black holes, we can usually elide switches and routers while verifying our supported set of invariants.

\vspace{0.1cm}
{\bf Scalability microbenchmark.}
Our system outperforms the alternative because it models the network as a set of parallel pipelines and checks each pipeline in isolation; but even for a single pipeline, policy size grows
as a function of the network size: for instance in our enterprise network example, the firewall has $O(n)$ ACL rules where $n$ is the number of hosts. Next, we consider the impact of policy
size on verification time.
\begin{figure*}[tb]
\subfigure[]{
    \centering
        \includegraphics[width=0.31\textwidth]{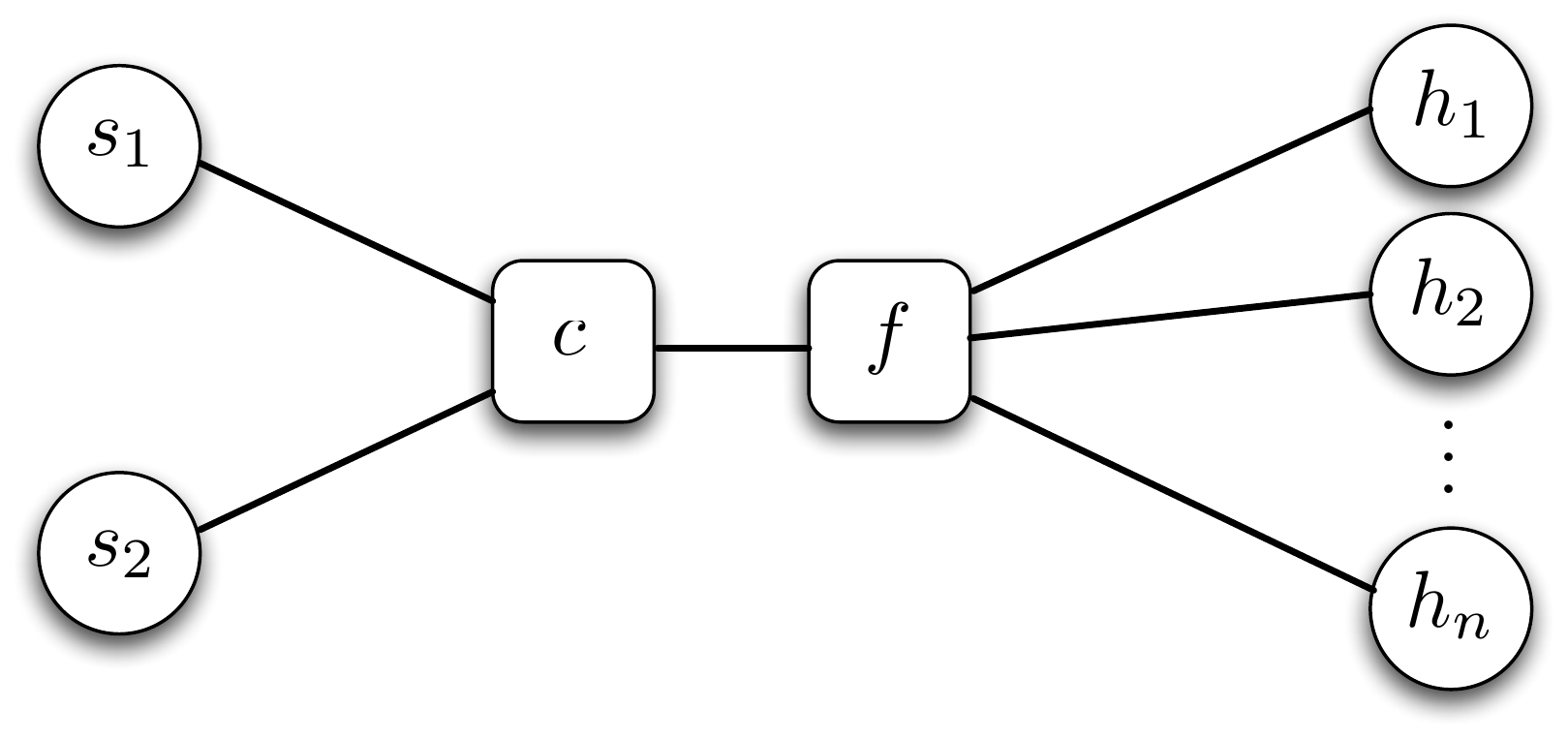}
        \label{fig:dataisolation}
}
\hfill
\subfigure[]{
  \centering
    \centering
    \includegraphics[width=0.31\textwidth]{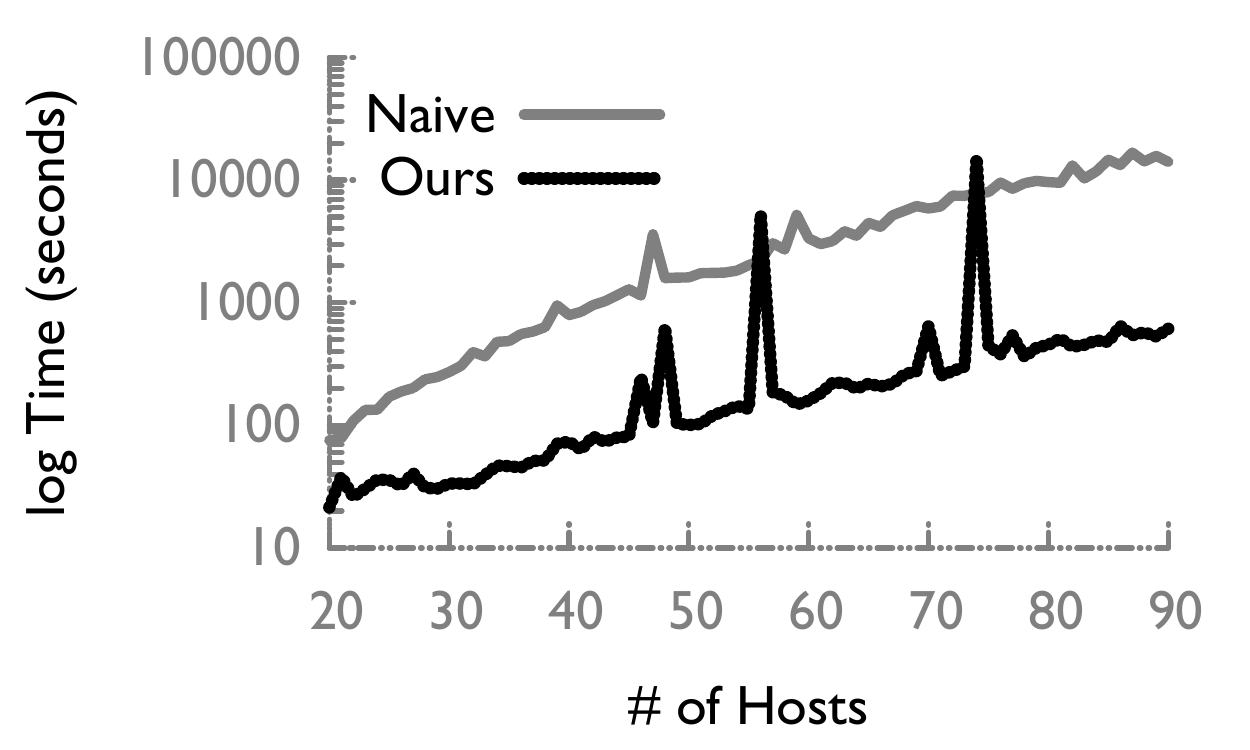}
    \label{fig:dataisoresults_s1}
}
\hfill
\subfigure[]{
\centering
    \centering
    \includegraphics[width=0.31\textwidth]{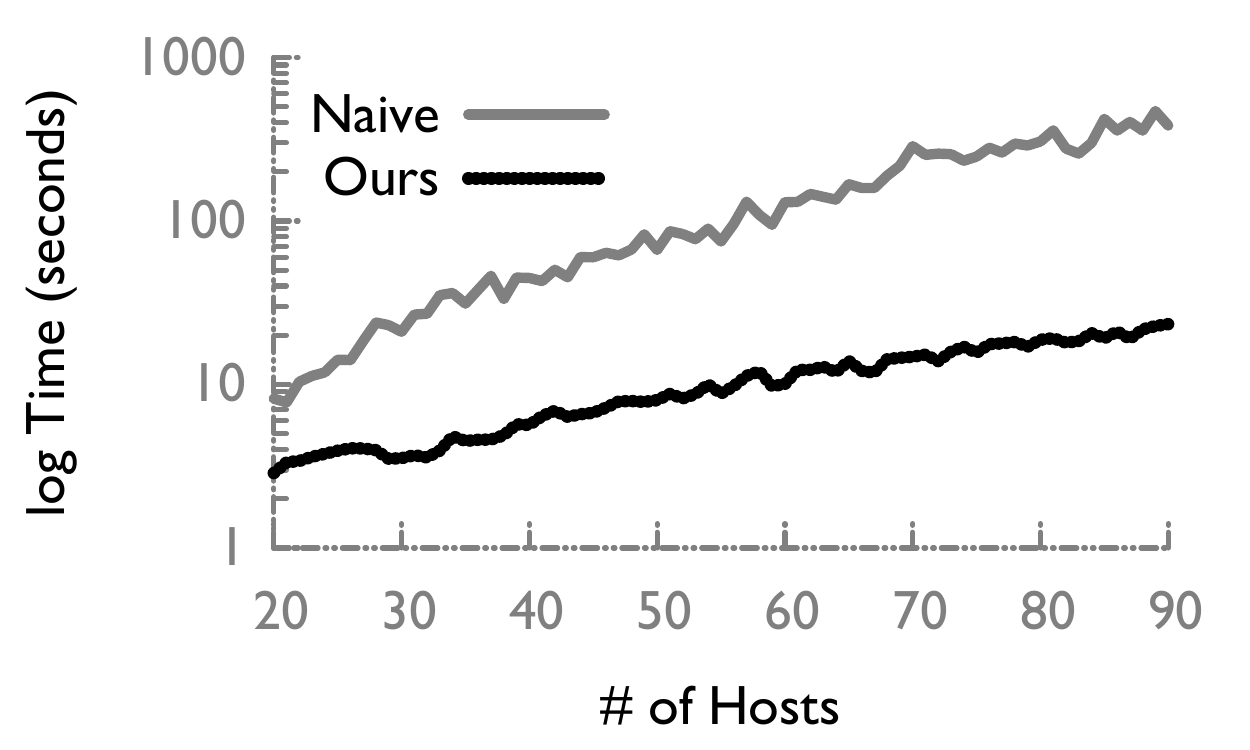}
    \label{fig:dataisoresults_s2}
}
\caption{\subref{fig:dataisolation} A content provider network with a content cache; \subref{fig:dataisoresults_s1} average pipeline verification time to check that hosts can access $s_1$ content; \subref{fig:dataisoresults_s2} average pipeline verification time to check that hosts cannot access $s_2$ content.}
\vspace{-0.2in}
\label{fig:dataisoall}
\end{figure*}
We study this in the topology in Figure~\ref{fig:policyscaling}, where two hosts, each with $n$ IP addresses, are separated by a firewall with $O(n^2)$ ACLs. We consider two invariants:
\begin{asparaenum}
\item Isolation: the two hosts can never communicate. For this, we install $n^2$ DENY exact-match rules in the firewall, corresponding to all pairs of the hosts' addresses.
\item Reachability: the two hosts can always communicate. For this, we install $n^2-1$ exact-match DENY rules in the firewall, blocking communication between all pairs of the hosts' addresses but one.
\end{asparaenum}
We measure how verification time increases with the number of firewall rules. Figure~\ref{fig:policyscale} shows the results. Note that while verification time does not grow rapidly, larger policy sizes result in larger models that need more memory to be expressed. At large enough sizes verification does not succeed because of memory pressure.

\subsection{Content Caches}
\label{sec:eval:cache}
Next, we consider the topology in Figure~\ref{fig:dataisolation}, where a set of hosts access a set of content servers placed behind a content cache and a firewall.
This setup is representative of content-provider networks, which employ content caches to improve the performance of user-facing services and reduce server load, but 
may place restrictions on what hosts can access a particular class of content.

A content cache typically implements an ACL to prevent clients from using caching to bypass policy: Consider Figure~\ref{fig:dataisolation} and suppose that each of the two content servers provides content to different clients. For example, when client $a$ requests ``http://xxx.com/latest-news,'' the request is served by $s_1$; when client $b$ requests the same name, the request is served by $s_2$. So, a request for the same name results in different content, depending on who is asking (a typical practice by content providers). To implement this policy, the provider configures the firewall to prevent communication between $a$ and $s_2$. However, every time $b$ accesses content from $s_2$, that content gets cached in the content cache and becomes available to $a$---unless the cache implements an ACL specifying that $a$ must not access content that originated in $s_2$. 

We use our system to verify two data-isolation invariants:
\begin{asparaenum}
\item Hosts $h_2 \ldots h_n$ can never access any content that originated  in server $s_2$.
\item Hosts $h_1 \ldots h_n$ can always access any content that originated in server $s_1$.
\end{asparaenum}
To implement these invariants, we configure the content cache with an exact-match rule denying access (to data that originated from server $s_2$) for each of hosts $h_2 \ldots h_n$. To make verification more challenging, we also configure the firewall with two exact-match rules denying access (to and from $s_2$) for each of these hosts. As above, we configure the middleboxes correctly, and our results are for the case where all invariants hold, but verification complexity is similar when the invariants are violated.

We measure the time taken to verify that the correct invariants hold for all hosts in the network. The content cache is flow-parallel: any content requests made by host $h_i$ do not affect the behavior of the content cache toward requests made by host $h_j$. The firewall is also flow-parallel and flow-preserving. Hence, a pipeline consisting of the content cache, the firewall and two end-hosts is RONO and can be checked in isolation. Figures~\ref{fig:dataisoresults_s1} and~\ref{fig:dataisoresults_s2} show pipeline verification time when we verify each pipeline in isolation (``Ours'') and when we verify the entire network (``Na\"ive''): in the case of $100$ accessing hosts, our system takes about $12$ minutes per pipeline, more than an order of magnitude faster than the na\"ive approach. Table~\ref{tab:largeverification} shows pipeline verification time for our system, when we have larger numbers of hosts: for a network of $1000$ hosts, it is a little above an hour; the na\"ive approach does not terminate in useful time.
Consistently with the results presented in Section~\ref{sec:eval:firewall}, verifying that certain hosts \emph{cannot} access certain content (Figure~\ref{fig:dataisoresults_s2}) is significantly faster than verifying that certain hosts \emph{can} access certain content (Figure~\ref{fig:dataisoresults_s1}).

\subsection{Generic Middleboxes}
\label{sec:eval:genmb}

\begin{figure*}[tb]
\centering
\subfigure[]{
\centering
        \centering
        \includegraphics[width=0.31\textwidth]{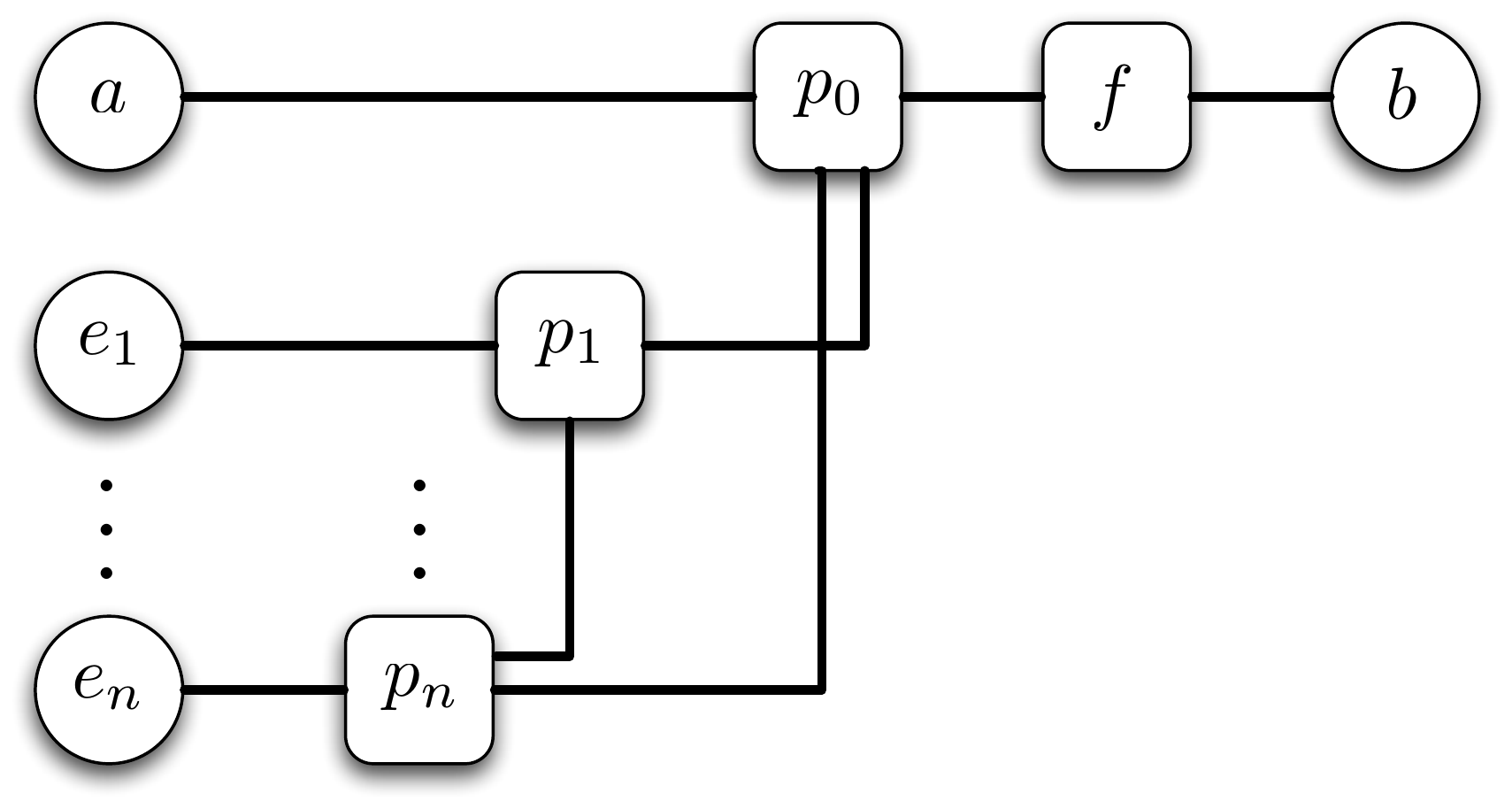}
        \label{fig:genericmbox}
}
\hspace{0.4in}
\subfigure[]{
\centering
    \centering
    \includegraphics[width=0.31\textwidth]{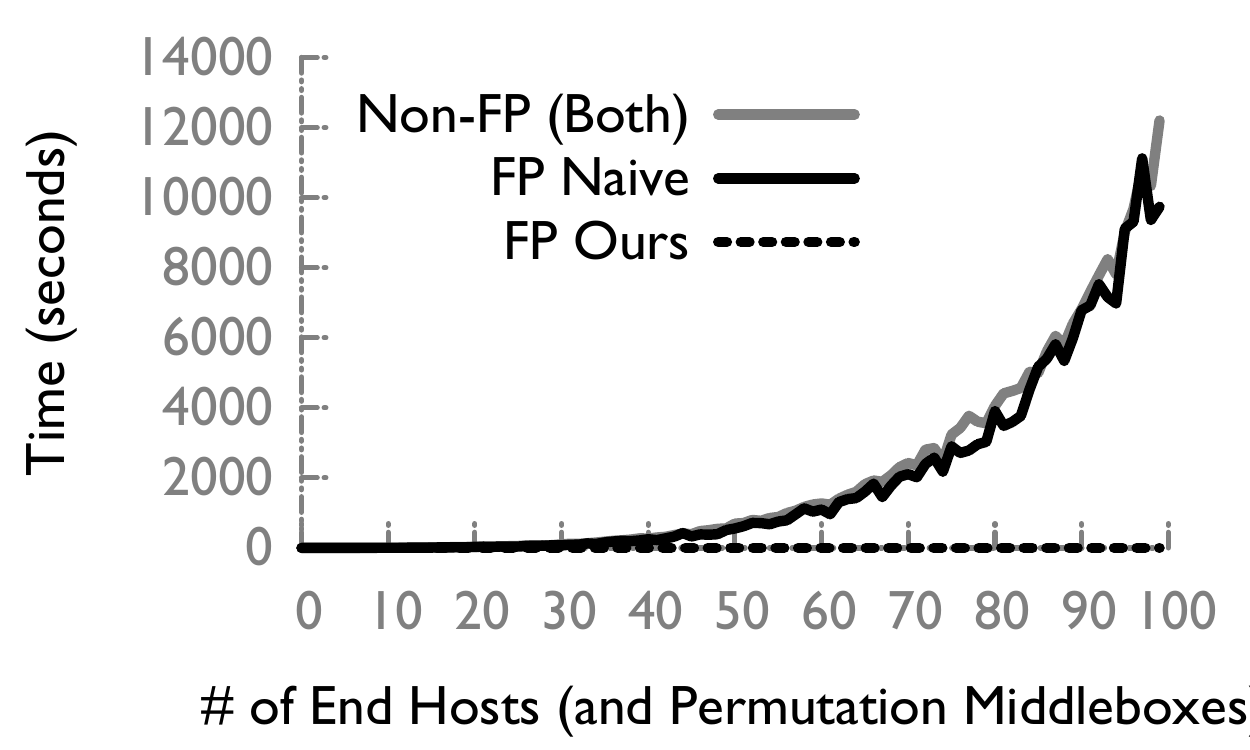}
    \label{fig:genmboxresult}
}
\caption{\subref{fig:genericmbox} Topology for evaluating generic middleboxes; \subref{fig:genmboxresult} time to verify node-isolation between a pair of hosts in the topology from Figure~\ref{fig:genericmbox}.}
\label{fig:genmball}
\vspace{-0.2in}
\end{figure*}

So far we have considered two existing, commonly used middleboxes; now we provide evidence that our system can handle other types of stateful middleboxes.

We consider the topology in Figure~\ref{fig:genericmbox}, where source $a$ is connected to destination $b$ through a ``permutation middlebox'' $p_0$ and a firewall $f$ that drops traffic between specific address pairs. Each of hosts $a$ and $b$ has multiple addresses; we denote $a$'s addresses by $a^i$ and $b$'s addresses by $b^j$. When $a$ sends a packet to $b$, $p_0$ replaces the packet's source and destination addresses with a different pair; how it chooses the new address pair depends on previously observed traffic from $a$ to $b$. For example, the box's configuration may dictate that: after observing a packet with source address $a^1$ and destination address $b^1$, in any future packet, source address $a^1$ will be replaced with $a^2$, and destination address $b^1$ will be replaced with $b^2$. Similarly, each permutation box $p_i$  permutes the source and destination addresses of packets from host $e_i$ to host $b$ based on previously observed traffic between these hosts. 

This setup is contrived, but it captures the complexity of any situation where a middlebox changes incoming packets based on traffic history. In our particular setup, the middlebox permutes the source and destination addresses of incoming packets, and its choice of new addresses determines whether the firewall will drop each resulting packet or not (potentially causing the violation of an isolation invariant). In a different setup, the middlebox might change some other part of incoming packets, and its choice would determine whether a downstream intrusion detection system would drop each resulting packet or not. 

We use our system to verify the following node-isolation invariant: host $a$ can never send traffic to host $b$. To implement this, we configure the firewall $f$ to drop all packets with $a$/$b$ address pairs. 
The permutation box $p_0$ is flow-parallel: its behavior with respect to $a$/$b$ traffic depends only on previously observed $a$/$b$ traffic, which allows us to check the $a$--$p_0$--$f$--$b$ pipeline in isolation. 
Figure~\ref{fig:genmboxresult} shows pipeline verification time as a function of the number of hosts (or permutation boxes, since we have one per host) in the network (consider only the ``Ours'' and ``Na\"ive'' curves, for the moment).
Our system takes a few tens of milliseconds per pipeline, independently from the number of nodes in the network; this is not the case for the na\"ive approach, which already takes two orders of magnitude longer for $20$ hosts.
Our system's pipeline verification time remains nearly constant with the number of hosts (and permutation boxes), because, in this particular setup, increasing the number of hosts does not affect the configuration size of $p_0$ or $f$ (the number of rules installed in the firewall depends on the number of $a$/$b$ address pairs). We do see a slight increase due to the fact that we allocate increasingly larger numbers of addresses and hosts. In a extreme test we
found that even with $30,000$ hosts and middleboxes our system could construct and verify the model in less than $5$
minutes.

\begin{figure*}[tb]
\centering
\subfigure[]{
        \centering
        \includegraphics[width=0.3\textwidth]{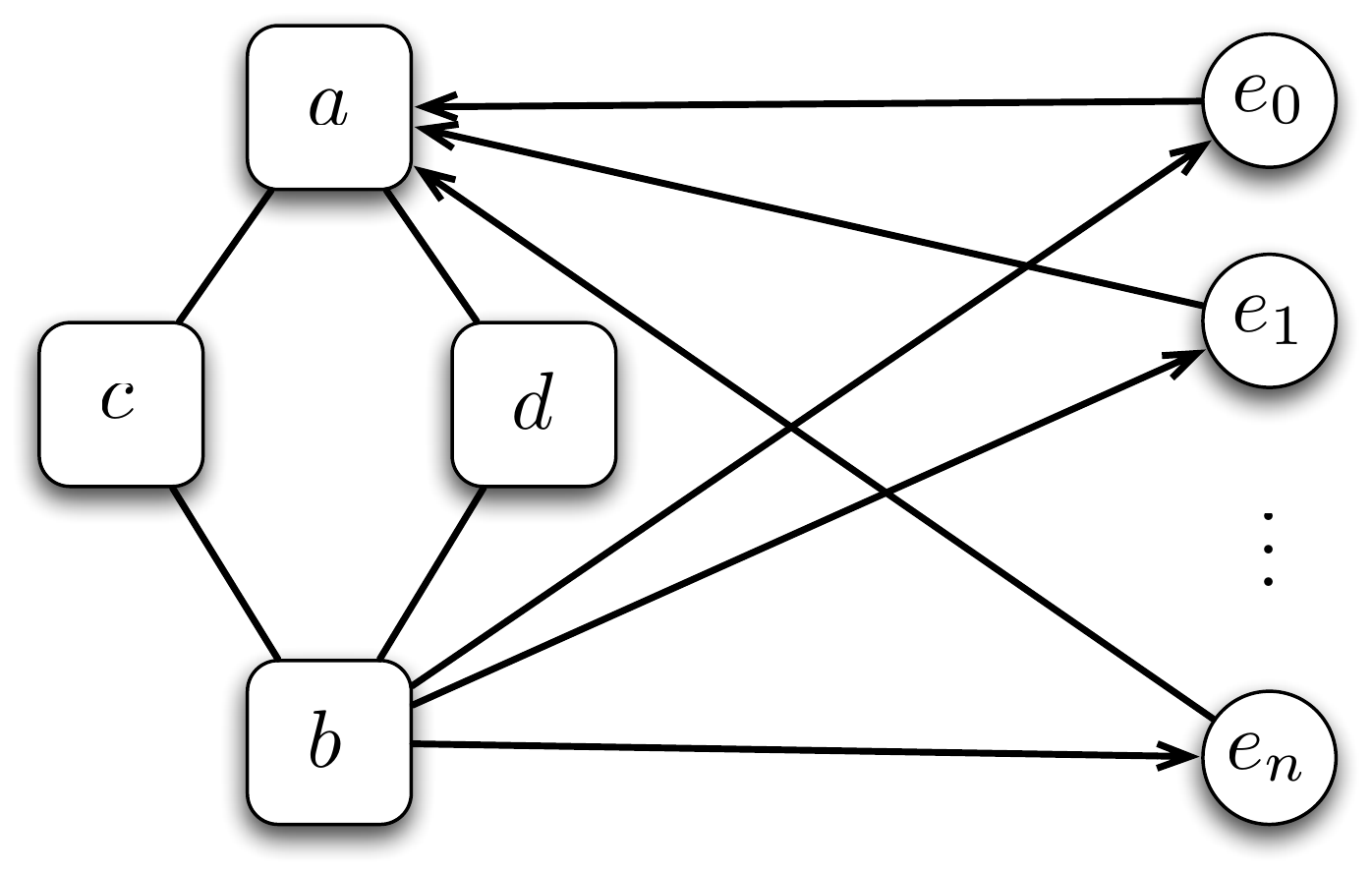}
        \label{fig:traversaltopo}
}
\hspace{0.4in}
\subfigure[]{
    \centering
    \includegraphics[width=0.3\textwidth]{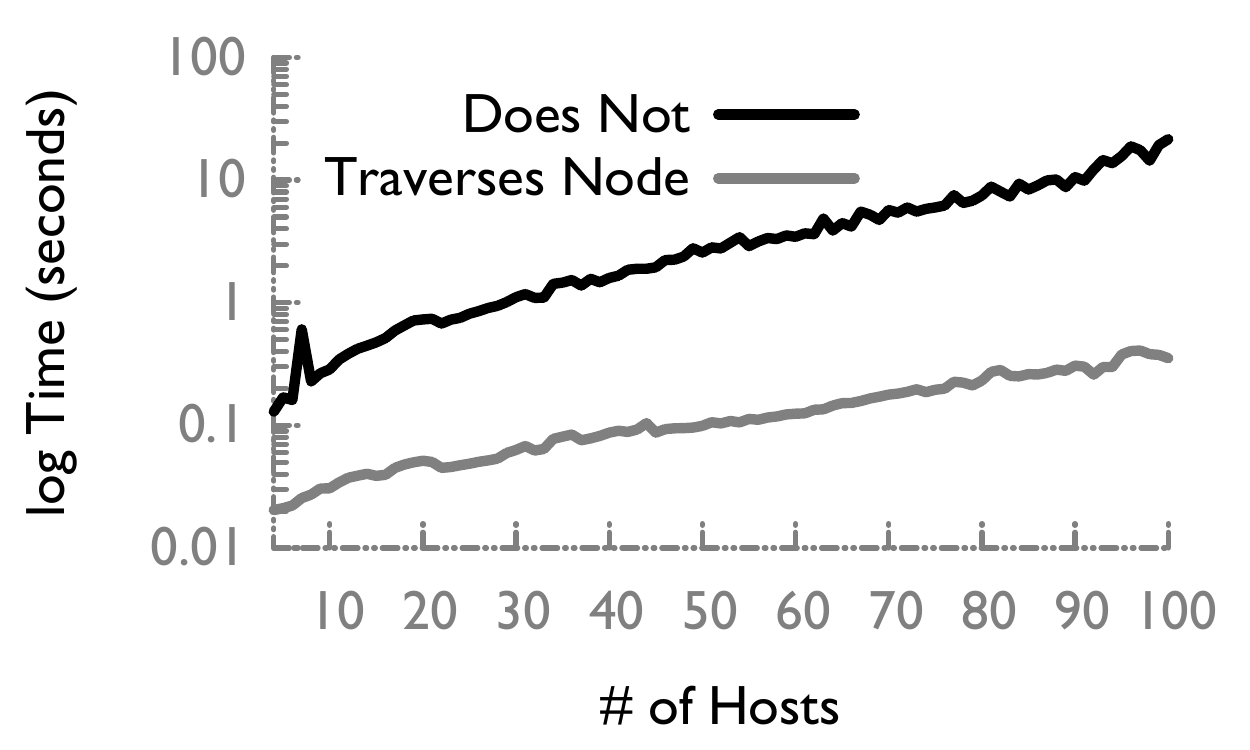}
    \label{fig:nodetrav}
}
\caption{\subref{fig:traversaltopo} Topology for evaluating node-traversal invariant; \subref{fig:nodetrav} verification time for node-traversal invariant.}
\vspace{-0.2in}
\label{fig:traversallall}
\end{figure*}

\subsection{The Benefit of RONO}
\label{sec:eval:nonrono}


So far we have considered only flow-parallel middleboxes, which our system was designed to handle efficiently; we now describe a scenario with non-flow-parallel middleboxes, where our system is as good as the na\"ive  approach.

We consider again the topology in Figure~\ref{fig:genericmbox} and the same node-isolation invariant as above (host $a$ can never send traffic to host $b$), and we configure the firewall to drop all packets with $a$/$b$ address pairs.
However, we make the permutation middleboxes non-flow-parallel: $p_0$ determines how to permute $a$/$b$ addresses based on previously observed traffic, not only from $a$ to $b$, but from any host $e_i$ to $b$. As a result, $p_0$'s behavior with respect to $a$/$b$ traffic depends on traffic previously carried by \emph{other} pipelines, and we cannot check the $a$--$p_0$--$f$--$b$ pipeline in isolation.  The ``Non-FP'' curve in Figure~\ref{fig:genmboxresult} shows pipeline verification time as a function of the number hosts in the network. 

The three curves in Figure~\ref{fig:genmboxresult} together show the benefit of RONO: when middleboxes are flow-parallel, our system leverages RONO and completes verification in tens of milliseconds (``Ours''); with the na\"ive approach (``Na\"ive''), verification takes as long as if the middleboxes were not flow-parallel (``Non-FP'').


\subsection{Node Traversal}
\label{sec:eval:extra}

We consider the topology in Figure~\ref{fig:traversaltopo}, where $a$, $b$, $c$ and $d$ are middleboxes that always forward all received packets: the forwarding tables are set up such that all traffic sent by any host ($e_0\ldots e_n$) is first sent to $a$, which depending on the destination forwards this traffic to either $c$ or $d$; finally, both $c$ and $d$ forward packets to $b$, which delivers them to the intended host. Note that checking node traversal requires considering the entire network (so RONO does not help us here). We use our system to verify a node-traversal invariant: that traffic from host $e_i$ to host $e_j$ always traverses middlebox $c$, for a given set of $\{i,j\}$ pairs. We configure the middleboxes such that the invariant holds for half the host pairs and not for the other half. Figure~\ref{fig:nodetrav} shows the average time to check this property for each host pair (averaged across the set of all considered host pairs).  


\section{Related Work}
\label{sec:related}
The earliest use of formal verification in networking focused on proving correctness and checking security
properties for protocols~\cite{clarke1998using, ritchey2000using}.
The first application of these techniques to control and data plane verification looked at verifying BGP configuration~\cite{feamster2004practical,
feamster2005detecting} in WANs.

\paragraphb{Verifying Forwarding Rules}
Recent efforts in network verification~\cite{MaiKACGK11,nsdi:CaniniVPKR12,kazemian2012header,khurshid2012veriflow,Verificare,FMACAD:SNM13,anderson2014netkat}
have focused on verifying the network dataplane by analyzing forwarding tables. Some of these tools including HSA~\cite{nsdi:KCZCMW13},
Libra~\cite{zeng2014libra} and VeriFlow~\cite{khurshid2012veriflow} have also developed algorithms to perform near real-time verification of simple 
properties such as loop-freedom and the absence of blackholes. While well suited for checking networks with static data planes they are
insufficient for dynamic datapaths.


\paragraph{Verifying Network Updates}
Another line of network verification research has focused on verification during configuration updates. This line of work can
be used to verify the consistency of routing tables generated by SDN controllers~\cite{KattaRW13,VanbeverRBFR13}.
Recent efforts~\cite{tr:MSRSegguru} have generalized these mechanisms and can be used to determine what parts of the
configuration are affected by an update, and verify invariants on this subset of the configuration. This line of work
has been restricted to analyzing policy updates performed by the control plane and does not address dynamic data plane elements
where state updates are more frequent and span a wider range.

\paragraphb{Verifying Network Applications}
Other work has looked at verifying the correctness of control and data plane applications. NICE~\cite{nsdi:CaniniVPKR12} proposed using static analysis to verify the
correctness of controller programs. Later extensions including~\cite{conext:uzniarPCVK12} have looked at improving the accuracy of NICE using
concolic testing~\cite{sen2006cute} at the cost of completeness. More recently, Vericon~\cite{ball2014vericon} has looked at sound verification of a restricted class of controllers.

Recent work~\cite{dobrescu2014software} has also looked at using symbolic execution to prove properties for programmable datapaths (middleboxes). This work in particular looked at verifying
bounded execution, crash freedom and that certain packets are filtered for stateless or simple stateful middleboxes written as pipelines and meeting certain criterion. The verification
technique does not scale to middleboxes like content caches which maintain arbitrary state.

\paragraphb{Finite State Model Checking}
Finite state model checking has been applied to check the correctness of many hardware and software based systems~\cite{book:CGD01}. Here the behavior
of a system is specified as a transition relation between finite state and a checker can verify that all reachable states from a starting configuration
are safe (\ie do not cause any invariant violation). Tools such as NICE~\cite{nsdi:CaniniVPKR12}, HSA~\cite{kazemian2012header} and others~\cite{CAV:SosnovichGN13}
rely on this technique. However these techniques scale exponentially with the number of states and for even moderately large problems one must chose between being
able to verify in reasonable time and completeness. Our use of SMT solvers allows us to reason about potentially infinite state and 
our choice of formulas are expressible in a way that guarantees termination of the SMT solver (\S\ref{sec:tractability:decidability}).

\paragraphb{Language Abstractions}
Several recent works in software-defined networking~\cite{IEEECOMM:FosterGRSFKMRRSWH13,sigcomm:VoellmyWYFH13,PLDI:GuhaRF13, FlowLog, NLOG} have
proposed the use of verification friendly languages for controllers. One could similarly extend this concept to provide a verification friendly
data plane language however our approach is orthogonal to such a development: we aim at proving network wide properties rather than properties for individual middleboxes.

Finally, parallel to our work Fayaz, et al.~\cite{fayaz2014buzz} use middlebox models to generate test packets to check
network invariants. Similar to us, they model middleboxes as state machines, however our models are expressed differently and they aim to test rather than verify networks.

\section{Conclusion}
\label{sec:conclusion}
On the one hand our work can be seen as merely a technical demonstration that one can verify isolation environments in large networks. 
However, our ambition is larger than merely providing operators with another verification tool. We hope that armed with the ability to 
verify both the static-datapath and dynamic-datapath aspects of a network, operators demand abstract middlebox models from
their vendors. This would (a) allow operators to enforce these aspects of the middleboxes, so that middlebox violations of prescribed
behavior can be detected and (b) allow operators to verify that their overall network meets their desired invariants. This would move networking 
from its current ad hoc practice to a more desirable state where invariants are explicitly expressed and rigorously enforced.




\bibliographystyle{abbrv}
\bibliography{main}
\end{document}